\documentclass[aps,prd,nofootinbib,twocolumn,longbibliography]{revtex4-1}
\usepackage[english]{babel}
\usepackage[autostyle, english = british]{csquotes}
\usepackage[CJKbookmarks, pdftex, bookmarksnumbered, bookmarksopen, colorlinks, citecolor=blue, linkcolor=blue]{hyperref}
\usepackage{amsmath,amssymb,mathtools}
\usepackage{graphicx}
\usepackage{appendix}
\usepackage{enumitem}
\setlist[enumerate]{topsep=6pt,parsep=-.5mm,leftmargin=5mm,}
 \MakeOuterQuote{"}
\def\be{\begin{equation}}
\def\ee{\end{equation}}
\newcommand*{\diff}{\mathop{}\!\mathrm{d}}

\begin{document}

\title{\large On the geometry of the black-to-white hole transition\texorpdfstring{\\}{} within a single asymptotic region}

\author{Muxin Han${}^{a,b}$, Carlo Rovelli${}^{c,d,e}$, Farshid Soltani${}^{f}$}

\affiliation{${}^a$ Physics Department, Florida Atlantic University, 777 Glades Road, Boca Raton FL 33431, USA.}

\affiliation{${}^b$ Department Physik, Institut f\"ur Quantengravitation, Theoretische Physik III, Friedrich-Alexander Universit\"at Erlangen-N\"urnberg, Staudtstr. 7/B2, 91058 Erlangen, Germany.}
\affiliation{${}^c$ Aix-Marseille University, Universit\'e de Toulon, CPT-CNRS, F-13288 Marseille, France.}
\affiliation{${}^d$ Perimeter Institute, 31 Caroline Street N, Waterloo ON, N2L2Y5, Canada} 
\affiliation{${}^e$ Department of Philosophy and the Rotman Institute of Philosophy, Western Ontario University,  London  N6A5B7, Canada}
\affiliation{${}^f$ Department of Physics and Astronomy, University of Western Ontario, London, ON N6A 3K7, Canada}

\begin{abstract} 

\noindent 
We write explicitly the complete Lorentzian metric of a singularity-free spacetime where a black hole transitions into a white hole located in its same asymptotic region. In particular, the metric interpolates between the black and white horizons. The metric satisfies the Einstein field equations up to the tunneling region. The matter giving rise to the black hole is described by the Oppenheimer-Snyder model, corrected with loop-quantum-cosmology techniques in the quantum region. The interior quantum geometry is fixed by a local Killing symmetry, broken at the horizon transition. At large scale, the geometry is determined by two parameters: the mass of the hole and the duration of the transition process. The latter is a global geometrical parameter.  We give the full metric outside the star in a single coordinate patch. 

\end{abstract}


\maketitle

\section{Introduction}

There is evidence in the sky of the presence of a huge number of black holes, with matter spiralling into them. General relativity predicts, arguably reliably, that this matter crosses the hole's horizon and reaches Planckian densities in a short proper time. What happens next is outside the reach of established physical theories. It involves the quantum behaviour of the gravitational field in the strong field domain.   

A possibility that has attracted interest \cite{Modesto:2004xx,Ashtekar:2005cj,Campiglia2008,Gambini2008a,Corichi2016,Olmedo2017,Ashtekar2018a,Ashtekar2018b,Munch2020,Zhang2020,Gan2020,Achour2020,Han2022,Han:2022rsx,Giesel2021,BarberoPerez2022,Husain2022,Husain2022a} is that the Einstein field equations are violated by a quantum tunnelling event, with a probability that depends on the curvature.  A natural scenario  is the black-to-white hole transition \cite{Rovelli2014,Haggard2014,DeLorenzo2016,Christodoulou2016,Bianchi2018b,DAmbrosio2021,Soltani2021,Rignon-Bret2022}, where the internal geometry of the hole undergoes a transition from trapped to anti-trapped (possibly through an intermediate non-trapped region) and the (outer) horizon tunnels from trapping to anti-trapping as well.   In this scenario the black hole evolves into a white-hole `remnant' living in the future of the parent black hole, in its same asymptotic region and location. Here we study the geometry of this process. 

We consider the case of a spherical black hole formed by the collapse of a homogeneous and pressure-less `star', as in the Oppenheimer-Snyder model \cite{OppenheimerSnyder}.  We disregard dissipative phenomena such as the Hawking radiation or the Perez dissipation into Planckian degrees of freedom \cite{Perez2015}. The inclusion of the former in the interior geometry of the black-to-white hole is studied in \cite{Rovelli2018a}. Dissipative phenomena are likely present in astrophysics and render the process irreversible. Here we only concentrate on the physics of the black-to-white transition alone, under the hypothesis that dissipative phenomena can be disregarded in a first approximation, as it can be done for a basketball bouncing on the floor.  The hypothesis is that the bounce can be described in a first approximation in terms of a few `large-scale' degrees of freedom.  We also neglect rotational degrees of freedom, but the causal structure of the spacetime we find has already similarities with the Kerr geometry, suggesting that rotation might not significantly alter the picture. 

We explore quantum effects only as local violations of the Einstein field equations, and not with a full quantum analysis.   We take one input from loop quantum gravity, following \cite{Kelly2021} and \cite{Giesel2022}: the correction to the Friedmann equation studied in loop quantum cosmology \cite{
Ashtekar2006,Yang2009,Agullo2013a,Assanioussi2018}. This same correction predicts a bounce at the end of the collapse of a homogeneous and pressure-less star, thus modifying the classical physics of the Oppenheimer-Snyder model. We match the exterior geometry to the star \cite{Munch2020,Achour2020}. As shown in \cite{Lewandowski2022}, the geometry of the interior of the hole outside the star is then uniquely determined by the evolution of the bouncing star and the local Killing symmetry. It turns out to be similar to the interior geometry of a Reissner–Nordstr\"{o}m black hole.  

We show that a quantum tunnelling briefly and locally violating the Einstein field equations around the horizon permits the bounce to happen also if no second asymptotic region exists. (See also \cite{Hergott2022}.)  The (surprising) compatibility of this scenario with the validity of the Einstein field equations outside the tunnelling region was pointed out in \cite{Haggard2014,Bianchi2018b}.     Crucially, we show that the horizon tunnelling region can be filled with an (effective) regular Lorentzian geometry. This geometry unravels the  possible global horizon structure of the black-to-white hole: there are no event nor global Killing horizons; there are only apparent horizons, and these keep the trapped and anti-trapped regions disconnected.  The metric we construct in this region is a proof of existence for a geometry with these features; as any trajectory in quantum tunnelling, it has no direct physical meaning. 

The geometry we found is consistent with previous general results. For instance, it belongs to the category A.I in the classification carried out in \cite{Carballo-Rubio}, the bounce of the star takes place in a non-trapped interior region $I$ bounded by two inner horizons, consistently with the analysis of matter collapse reported in~\cite{Achour2020}, and the exterior geometry fails to be exactly static in the vicinity of the horizon at the transition~\cite{Schmitz2021}.

Outside the star, the geometry we find depends on parameters that have transparent physical meaning. Two of them are measurable from a distance: in natural units, they are the mass $m$ of the star, and the duration ${\cal T}$ of the full process, from the collapse of the star into its black horizon to its emersion from the white horizon. Other parameters do not affect the large scale geometry; some of them may be measured locally around the horizon tunnelling region: they determine its size. Interestingly, ${\cal T}$ is a global geometric parameter (like the radius of a cylinder), not determined by the local geometry outside the quantum tunnelling region.    A quantum theory of gravity should determine the values, or the probability distribution, of all parameters.  Steps in this direction have been taken in \cite{Christodoulou2016,DAmbrosio2021,Soltani2021}.

Section~\ref{s:star} deals with the physics of the bounce of the collapsing star. This was called ``region $\mathcal C$'' in \cite{DAmbrosio2021}. Section~\ref{s:ext} deals with the physics of the interior of the black hole where the curvature reaches Planckian value. This was called ``region $\mathcal A$'' in \cite{DAmbrosio2021}. Section~\ref{s:B}
deals with the physics of the horizon tunneling region. This was called ``region $\mathcal B$'' in \cite{DAmbrosio2021}. Different physical processes happen in the three regions, and they must be dealt with separately. In Section~\ref{largescale} we describe the physical meaning and the large scale geometry of the spacetime we have built. Global coordinates for this spacetime are given in Section~\ref{globalcoordiates}. In Section~\ref{effective} we build a Lorentzian metric for the $\mathcal B$ region and in Section~\ref{horizons} we study its horizon structure.
 
\section{The star}
\label{s:star}

The metric inside a spherical pressure-less star of uniform density $\rho$ and total mass $m$ can be written in comoving coordinates $(T,R)$ as 
\be
\diff s^2=-\diff T^2+a^2(T)(\diff R^2+R^2\diff\Omega^2)\, ,
\label{star}
\ee
where $\diff \Omega^2$ is the metric of the unit 2-sphere, $R\in[0,R_{boundary}]$ and $a(T)$ is known as the scale factor. The radial comoving coordinate of the boundary of the star can be chosen to be $R_{boundary}=1$ without loss of generality. The uniform density of the star is then $\rho=m/\frac43\pi a^3$.  

\begin{figure}[t]
\begin{center}
\includegraphics[width=6cm]{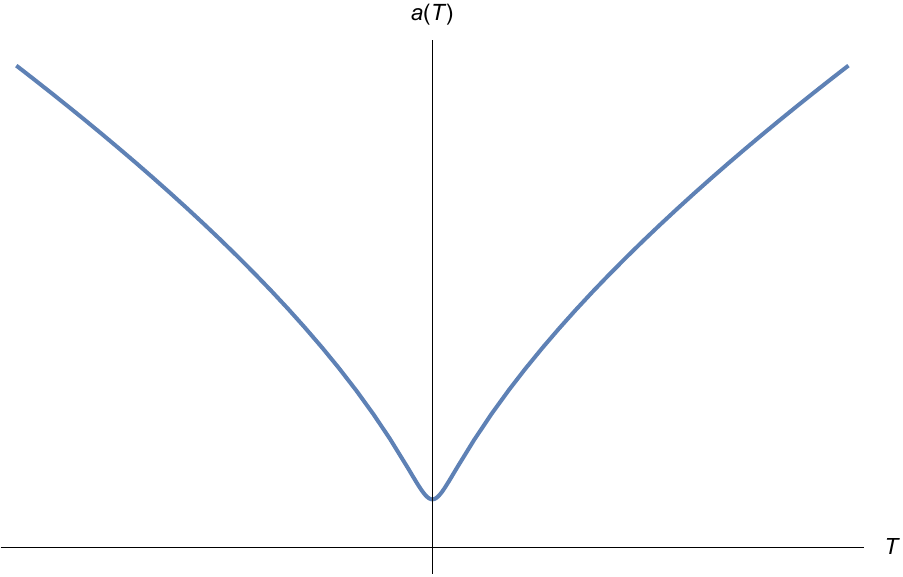}
\caption{The scale factor $a(T)$ in eq.~\eqref{aT} that gives the standard LQC bounce.}
\label{bounce2}
\end{center}
\end{figure}

The Einstein field equations imply that $a(T)$ satisfies the Friedmann equation.   
Loop quantum gravity adds a quantum correction term to this equation \cite{Agullo2013a}, which becomes 
\be
\frac{\dot a^2}{a^2}=\frac{8\pi G}3 \rho \left( 1-\frac\rho{\rho_{c}} \right),
\ee
where the critical density $\rho_{c}= \sqrt{3}c^2/(32 \pi^2 \gamma^3 \hbar G^2)\sim c^2/\hbar G^2$, $\gamma$ being the Barbero-Immirzi parameter, is a constant with the dimension of a density and Planckian value. Equivalently, defining a constant $A=3/(2\pi\rho_{c})$, and using units in which $G=c=1$ from now on, we can write
\be
\dot a^2=\frac{2m}{a}-\frac{Am^2}{a^4}.
\label{eqaT}
\ee
In these units, the constant $A\sim \hbar \sim m_{Pl}^2$ has dimension of a squared mass and Planckian value. The last equation can be integrated, giving 
\be
a(T)=\left(\frac{9 m T^2+Am}2\right)^{1/3}\,.
\label{aT}
\ee
As shown in Fig.~\ref{bounce2}, $a(T)$ is positive for the whole range $T\in [-\infty, \infty]$: it decreases for $T<0$, reaches a minimum $a_0=\sqrt[3]{Am/2}$ for $T=0$ and then increases for $T>0$. This is the characteristic bounce of loop quantum cosmology. This feature of $a(T)$ assures that the line element in eq.~\eqref{star} is well defined everywhere.

The coordinate $T$ is the proper time along the comoving worldlines, hence it is also the proper time on the boundary of the star. This means that eqs.~(\ref{eqaT}-\ref{aT}) give the evolution of the physical radius $r_b(T)=a(T)R_{boundary}=a(T)$ of the star in its own proper time, hence
\be
\dot r_b^2=\frac{2m}{r_b}-\frac{Am^2}{r_b^4}\,. 
\label{boundaryint}
\ee

\section{The exterior}
\label{s:ext}

Where the quantum corrections are negligible, an exact solution of the Einstein field equations is given by the geometry for the star described above (with negligible $A$) surrounded by the Schwarzschild geometry. The Schwarzschild geometry (i) matches the geometry of the star on the star's surface  \cite{OppenheimerSnyder}, (ii) is spherically symmetric, and (iii) is characterised by a killing field in addition to those related to the spherical symmetry. (This is timelike outside the horizon, where it enforces  the stationarity of the exterior geometry, and spacelike inside the horizon, where the geometry is \emph{not} stationary.) In \cite{Lewandowski2022}, it is shown that if we do include the quantum corrections, that is $A\ne 0$, these three features are realized by the metric 
\be
\diff s^2=-F(r)\diff t^2+\frac{\diff r^2}{F(r)}+r^2\diff \Omega^2\, ,
\label{ext}
\ee
where 
\be
F(r)=1-\frac{2m}r+\frac{Am^2}{r^4}\,.
\label{F}
\ee 
This geometry is clearly spherically symmetric and admits the killing field $\xi=\partial_t$.

A thin shell freely falling in it has the conserved quantity $E=F(r)\dot t$, where $E\sim1$ if the shell starts with vanishing velocity at large distance. The normalization of its proper time gives   
\be
-1=-\frac{1}{F(r)}+\frac{\dot r^2}{F(r)},
\ee
from which it follows that 
\be
\dot r^2=\frac{2m}r-\frac{Am^2}{r^4},
\label{boundaryext}
\ee
which is exactly eq.~\eqref{boundaryint} (as it should be, since this equation gives the evolution of the physical radius $r$ of the shell in its own proper time). This shows that the surface of the pressure-less star is in free fall in this metric.

The exterior geometry depends on two parameters: the total mass $m$ of the star and the constant $A\sim m^2_{Pl}$ characterizing the quantum correction to the Friedmann equation. If $m\gg m_{Pl}$, the last term in eq.~\eqref{F} gives a negligible correction to the Schwarzschild geometry for $r$ of order $m$ or larger.   

Interestingly enough, the same exterior metric can  be derived by starting from Schwarzschild spacetime and  considering quantum corrections coming from loop quantum gravity \cite{Kelly:2020uwj}.


\begin{figure}[t]
\begin{center}
\centerline{\includegraphics[width=4.5cm]{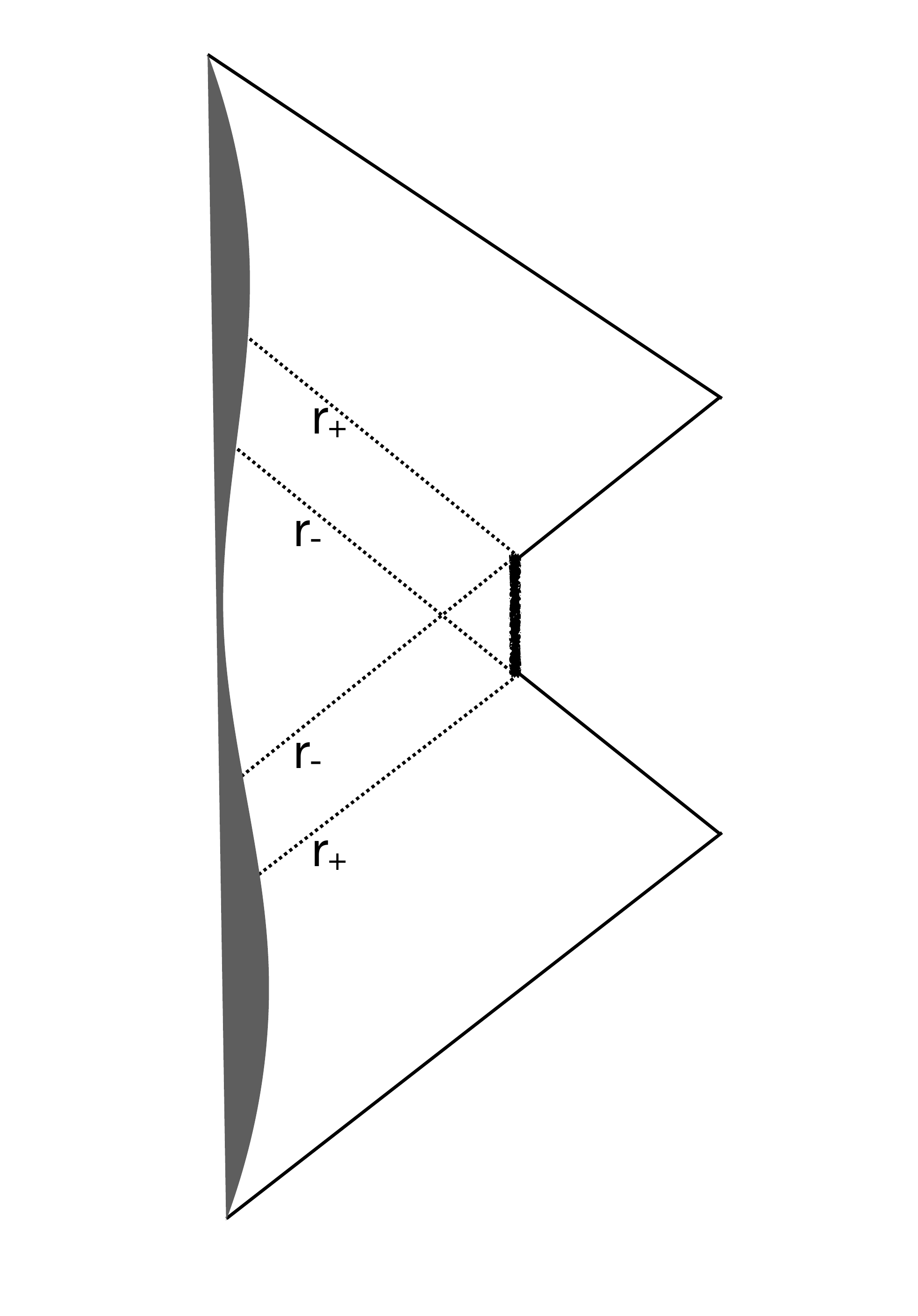}\hspace{.5cm} \raisebox{4mm}{\includegraphics[width=3.1cm]{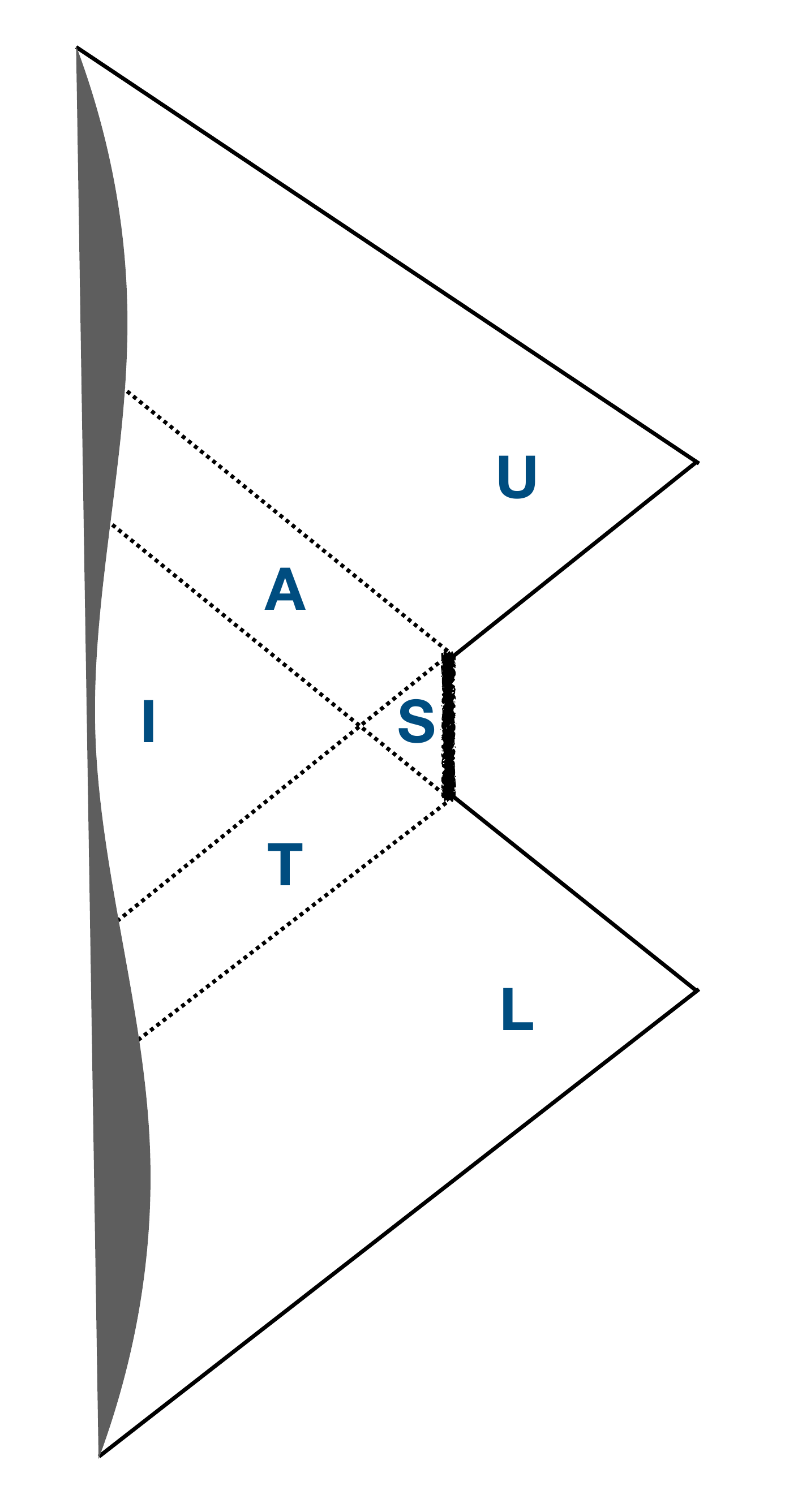}}}
\caption{Conformal diagram of the maximal extension of the spacetime representing the star and the exterior region defined by eqs.~(\ref{ext}-\ref{F}).}
\label{regions}
\end{center}
\end{figure}

Let us  study this geometry. Killing horizons are defined by the vanishing of the norm of the killing field  $\xi=\partial_t$, namely by  $g_{tt}=-F(r)=0$.    The investigation of the roots of $F(r)$, which is thoroughly performed in Appendix A, shows that there are two real roots $r_\pm$, see eq.~\eqref{solperturb}, and thus two killing horizons.   For $m\gg m_{Pl}$, that is $m^2\gg A$, 
\be
r_+ = 2m + O(A/m) \sim r_{Schwarzschild}
\ee
is the outer horizon of the black hole and it is located in the classical region, while 
\be
r_- = \sqrt[3]{Am/2} + O(A^{2/3}/m^{1/3}) \sim \sqrt[3]{m/m_{Pl}}\,l_{Pl}
\ee
is an inner horizon and it is located  inside the quantum region, that is the region where the spacetime curvature has Planckian size. A direct study of the metric in eqs.~(\ref{ext}-\ref{F}) shows that $r=r_\pm$ are also apparent horizons. That is, they separate trapped, non-trapped and anti-trapped regions.

Studying the geodesics of the spacetime it is easy to see that the coordinate $t$ diverges on all these horizons. The metric, however, is regular on them and it can be extended past them. This extension follows closely the extension of the Reissner–Nordström metric and it will be performed shortly. The spacetime resulting from the maximal extension of the metric is represented in the Penrose diagram in Fig.~\ref{regions}.    The spacetime comprises of several regions separated by the horizons:
\begin{itemize}
\item There are two asymptotic regions, a ``lower'' region $L$ bounded by a lower outer horizon and an ``upper'' region $U$ bounded by an upper outer horizon, where $r> r_+$.
\item There are a trapped region $T$ and an anti-trapped region $A$ where  $r_-< r < r_+$.
\item There are two interior non-trapped regions; one inner region $I$ next to the star's bounce where $r_{b}(\tau)<r<r_-$, $r_{b}(\tau)$ being the wordline of the star's boundary satisfying eq.~\eqref{boundaryext}, and an interior region $S$ bounded by a timelike singularity where $0<r<r_-$.
\end{itemize}
The bounce of the star takes place in the non-trapped interior region $I$ bounded by the two inner horizons. As mentioned, this is consistent with the analysis of matter collapse  in~\cite{Achour2020}.

The coordinate system $(t,r)$ separately cover each of the six regions represented in Fig.~\ref{regions}. In order to maximally extend the metric in eqs.~(\ref{ext}-\ref{F}) we can proceed as follows. The metric can be trivially rewritten as
\be
\diff s^2=F(r)\left(-\diff t^2+\frac{\diff r^2}{F^2(r)}\right)+r^2\diff \Omega^2,
\label{ext2}
\ee
which suggests to introduce a generalized tortoise coordinate $r_*$ satisfying 
\be
\diff r_*=\frac{\diff r}{F(r)}.
\label{r*r}
\ee
The integration of this differential equation is performed in Appendix B and the analytical expression of $r_*(r)$ can be found in eq.~\eqref{r*exact}. The function $r_*(r)$ is separately well-defined in each of the six regions represented in Fig.~\ref{regions}, but it diverges logarithmically on the horizons. By substituting eq.~\eqref{r*r} in eq.~\eqref{ext2} we get
\be
\diff s^2=F(r(r_*))\left(-\diff t^2+\diff r_*^2\right)+r^2\diff \Omega^2,
\ee
which allows us to introduce the null coordinates 
\begin{eqnarray}
&&u=r_*(r)-t, \label{uv} \\ 
&&{v=r_*(r)+t}, \label{vu}
\end{eqnarray}
in terms of which the metric reads
\be
\diff s^2=F(r(u,v))\,\diff u\,\diff v+r^2(u,v)\diff \Omega^2.
\label{null}
\ee
The function $r(u,v)$ is implicitly defined by 
\be
2r_*(r)=v+u.
\label{r*vu}
\ee
The sign of the coordinate $u$ defined here is the inverse of the one normally used in the literature.  This convention much simplifies later formulas. 

The new coordinates $u$ and $v$ diverge respectively on the two upper horizons and the two lower horizons, so the coordinate system $(u,v)$ is still ill defined on every horizon, thus preventing any extension of the spacetime. We can however use the coordinate system $(v,r)$, whose metric reads 
\be
\diff s^2=-F(r)\diff v^2+2\diff v\diff r+r^2 \diff \Omega^2,
\label{rv}
\ee
to cover in a single patch either regions $L,\, T,\, I$ or regions $S, \, A$ or region $U$, and the coordinate system $(u,r)$, whose metric reads
\be
\diff s^2=-F(r)\diff u^2+2\diff u\diff r+r^2 \diff \Omega^2,
\label{ru}
\ee
to cover in a single patch either regions $U,\, A,\, I$ or regions $S, \, T$ or region $L$. This allows all these regions to be glued as in Fig.~\ref{regions} and shows that they define together the maximal extension of the spacetime.

It is convenient to choose $v=0$ as the advanced time in which the star's boundary enters the lower outer horizon $r_+$ and $u=0$ as the retarded time in which the star's boundary exits the upper outer horizon $r_+$.   That is: the origin of the advanced time in $L$ is determined by the moment the star collapses into its own outer horizon forming a black hole and the origin of the retarded time in $U$ is determined by the moment the star emerges from its own outer horizon ending the white hole. 

There is a subtle but important fact to consider.  The function $r_*(r)$ does not enter the definition of the metrics of the two patches in eqs.~(\ref{rv}-\ref{ru}). However, it enters the coordinate transformation on the overlap:
\be
u=2r_*(r) -v.
\ee
The integral $r_*(r)$ of eq.~\eqref{r*r} depends on an integration constant which can be fixed by selecting $r_*$ at some location (see eq.~\eqref{r*exact}).   The integral however diverges on the two (real) zeros of $F(r)$, namely on the two horizons.   Hence so does $r_*$.  We can therefore define $r_*(r)$ in different patches across horizons, but we must remember that doing so adds a distinct constant in each patch. That is, $r$ is defined globally, but $r_*(r)$ is defined up to a constant in each patch. This will play a key role below.


\section{The horizon tunneling}
\label{s:B}

The spacetime represented in Fig.~\ref{regions} cannot be a realistic approximation of the dynamics of a black hole, because as soon as the Hawking evaporation process is taken into account, the lifetime of the black hole as seen from the lower asymptotic region $L$ becomes finite. This is incompatible with the geometry of  Fig.~\ref{regions}, where this lifetime is infinite. 

The dynamics of the horizon at the end of the evaporation is governed by  quantum gravity. Here, following~\cite{Haggard2014,DeLorenzo2016,Christodoulou2016,Bianchi2018b,DAmbrosio2021,Soltani2021,Rignon-Bret2022}, we consider the possibility that there is a non-vanishing probability for the geometry around the black hole  horizon to tunnel into the geometry around white hole horizons, via a local process within a single asymptotic region.

We do not compute the probability for this transition (for steps in this direction, see \cite{Christodoulou2016,DAmbrosio2021,Soltani2021}.)
Analogy with non-relativistic quantum tunnelling suggests that the tunnelling probability could be of the order of $\exp\{-m^2/\hbar G\}=\exp\{-m^2/m_{Plank}^2\}$. If so, the transition probability is suppressed until the very last phases of the evaporation, where $m\sim m_{Planck}$, and the tunneling physics we specify below describes the tunneling geometry at the end of the evaporation. If instead the transition probability is not so suppressed at larger $m$, the tunnelling may happen earlier (an heuristic argument in favor of a shorter timescale is given in \cite{Haggard2014,Haggard20162}). 

Notice however that even if we entirely disregard the Hawking radiation and the consequent decrease of $m$ with time, any nonzero transition probability implies \emph{anyway} that sooner or later the tunnelling happens, because small probabilities pile up with time, as in ordinary radioactivity. Thus the inclusion of the evaporation process in the analysis should not alter the resulting qualitative picture. The 
tunnelling we describe below can happen in any case, unless it is forbidden by something that at present we cannot see.  Hence below we neglect the Hawking radiation and we make no assumption about the transition amplitude, which can be arbitrary small. We will see below which parameter of the resulting geometry depends on this quantum transition amplitude. 


In this section we construct the spacetime describing the horizon tunnelling. We do so starting from the maximally extended spacetime in Fig.~\ref{regions},  cutting away a part of it, inserting a new spacetime region and gluing some resulting boundaries.   We start by excising a part of the maximally extended spacetime described above.

Fix three constants $r_\alpha,r_\beta$ and $r_\delta$ with the dimension of a length and satisfying $r_\alpha<r_-<r_+<r_\delta<r_\beta$.  We shall also use $\delta\equiv r_\delta-r_+>0$. The geometry we are going to define is thus based on these four parameters in natural units: $m,r_\alpha,r_\beta,r_\delta$ (plus $A\sim m^2_{Pl}=\hbar G$ that determines a scale). We are particularly interested in the regime where $r_\alpha$ is close to $r_-$, and $r_\beta$ (and so $r_\delta$) is close to $r_+$.

\begin{figure}[t]
\begin{center}
\centerline{\includegraphics[width=4.5cm]{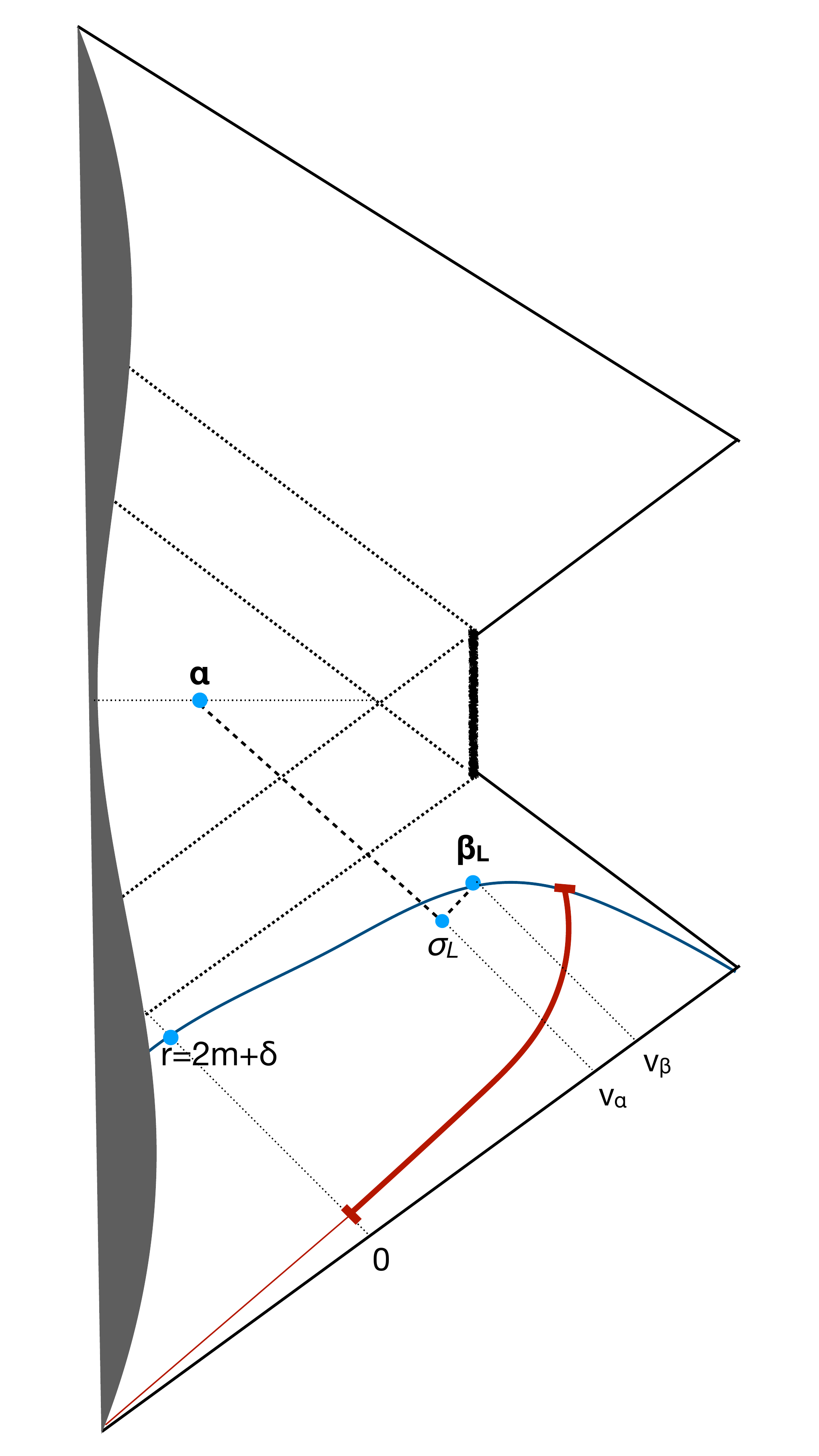}}
\caption{The points (2-spheres) $\alpha$, $\beta_L$, $\sigma_L$. In blue, the surface $t=0$ and its intersection with $v=0$. In red, the worldline of an observer at a constant distance.}
\label{all}
\end{center}
\end{figure}

\begin{figure}[t]
\begin{center}
\centerline{\includegraphics[width=4.7cm]{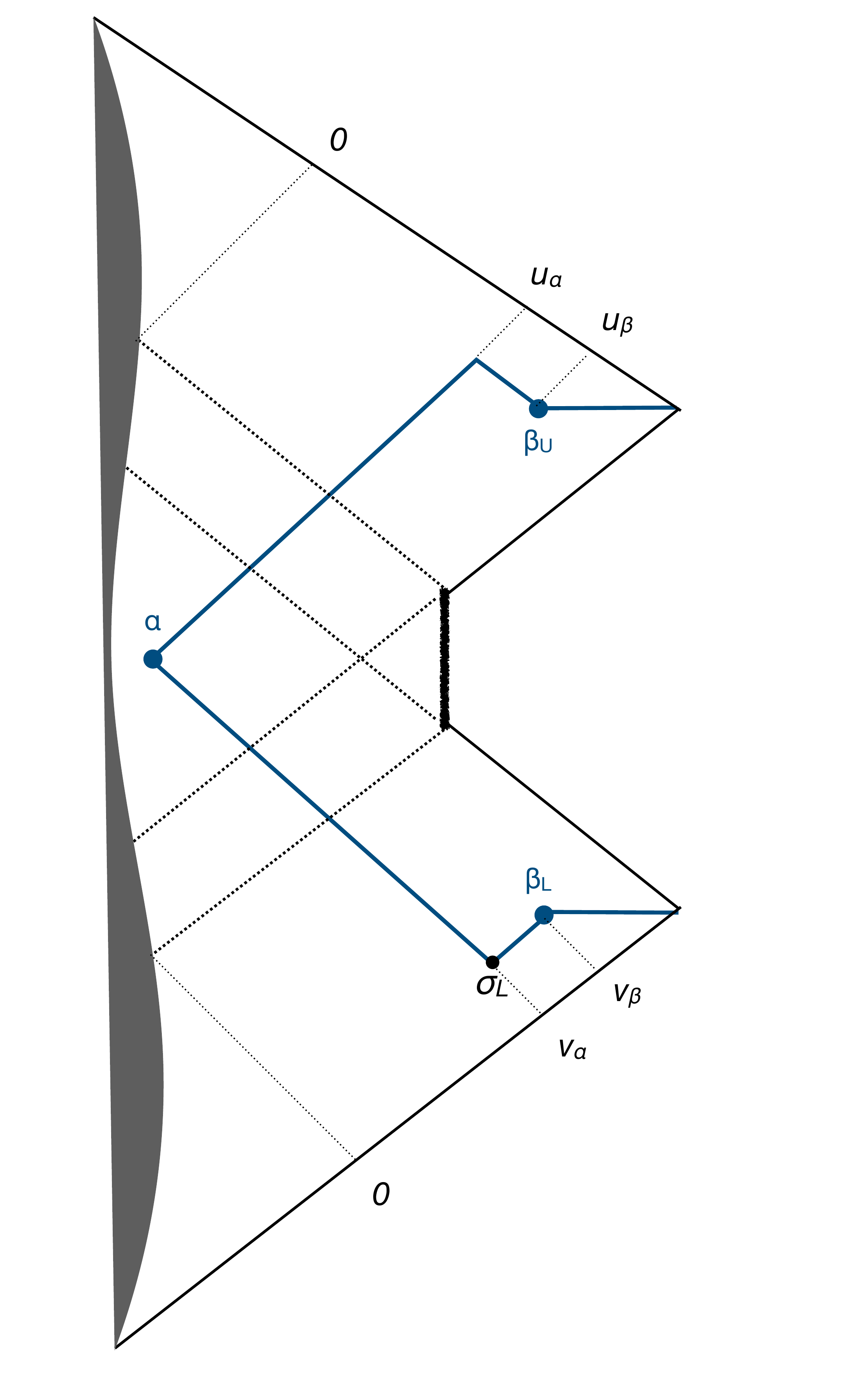}}
\caption{The blue line is the boundary of the region that is excised, because it is not a good approximation of the spacetime of a physical black hole. The two horizontal portions of the blue line are identified; the excised region is replaced by a non-singular geometry.}
\label{bouncestarB}
\end{center}
\end{figure}

In region $I$, consider the $t=constant$ surface containing the bounce point of the star (see Fig.~\ref{all}). On this surface, let $\mathbf \alpha$ be the point with radial coordinate $r_{\alpha}$ (the first of the parameters for the geometry we are constructing). Let $v_\alpha$ be the advanced time of $\mathbf \alpha$. This is going to be the advanced time at which the horizon transition begins.

It is a simple exercise to express $v_\alpha$ as a function of $r_{\alpha}$. First, we have to determine the advanced time $v_{b}$ of the bounce point of the star. This can be determined from a standard calculation in general relativity and it is of order $m$.   The $t$ coordinate of the star's bounce is then, from eq.~\eqref{vu},
\be
t=v_b-r_*(r_{b}),
\ee
and since $\alpha$ is on the same $t$ surface, we also have
\be
t={v_\alpha}-r_*(r_{\alpha}).
\ee
The two relations imply
\be
{v_\alpha}=v_b+r_*(r_{\alpha})-r_*(r_{b}),
\ee
which does not depend on the undetermined integration constant of $r_*(r)$ in $I$. If $r_\alpha$ approaches $r_-$, the advanced time ${v_\alpha}$ can be arbitrarily long, as $r_*$ diverges in $r_-$.  We are particularly interested in this regime, where the time from the collapse of the star to the onset of the horizon tunnelling can be arbitrarily long. The radial coordinate $r_{\alpha}$ is going to be the maximum radius on the $t=constant$ surface in region $I$ for which the metric constructed in section~\ref{s:ext} is a good approximation of the spacetime of a black hole.

Next, observe that all constant-$t$ time surfaces in the $L$ region intersect the line $v=0$ \emph{outside the outer horizon}. (Recall that $v=0$ is the advanced time of the point where the boundary of the star enters the outer horizon.) We insist on this detail because it is a counter-intuitive feature of classical general relativity.  The later the time $t$, the closer to the horizon the constant-$t$ surface intersects $v=0$.  Consider the constant time surface that intersect $v=0$ at the radius $r_\delta=r_++\delta$ (the second of the parameters that we introduce). An arbitrarily small $\delta$ determines an arbitrarily late $t$.  (Later on, this time $t$ will determine the reflection surface under time inversion.)

Without loss of generality, we can call this surface $t=0$, because this simply amounts to fixing once and for all the integration constant of $r_*(r)$ in region $L$, and represent it in a conformal diagram as in Fig.~\ref{bouncestarB}. Explicitly, the intersection has coordinates $v=t=0$ and $r\sim 2m+\delta$. Therefore eq.~\eqref{vu} fixes $r_*(r_\delta)=0$.    
%

Consider then the point $\beta_L$ with radius $r_\beta$ (the third parameter we introduce) on the $t=0$ surface. Let $v_\beta$ be its advanced time.  We assume that the constants we have introduced are such that $v_\beta> v_\alpha$. (Given $r_\alpha$ and $r_\beta$, this is always possible by taking $\delta$ small enough).  We are particularly interested in the regime in which $r_\beta$ is close to $r_+$. Since $r_\beta>r_\delta=2m+\delta>r_+$, this means that $\delta$ must be small.   Let  $\sigma_L$ be the intersection of the past outgoing null geodesic originating in $\beta_L$ and the past ingoing null geodesic originating in $\alpha$. These null geodesics are represented as dashed lines in Fig.~\ref{all} and as blue lines in Fig.~\ref{bouncestarB}.

The above construction in the regions $L,T,I$ can be repeated  symmetrically in the upper regions $U,A,I$. See Fig.~\ref{bouncestarB}.   By symmetry, the retarded time coordinate $u$ of $\alpha$ in the upper region is $u_\alpha=v_\alpha$. We consider a constant-$t$ surface in the upper region $U$ as well, which we can call $t=0$ by fixing the integration constant of $r_*(r)$ in region $U$, and a point $\beta_U$ with radius $r_{\beta}$. Its retarded time is $u_\beta=v_\beta$.

\begin{figure}[t]
\begin{center}
\centerline{\includegraphics[width=5cm]{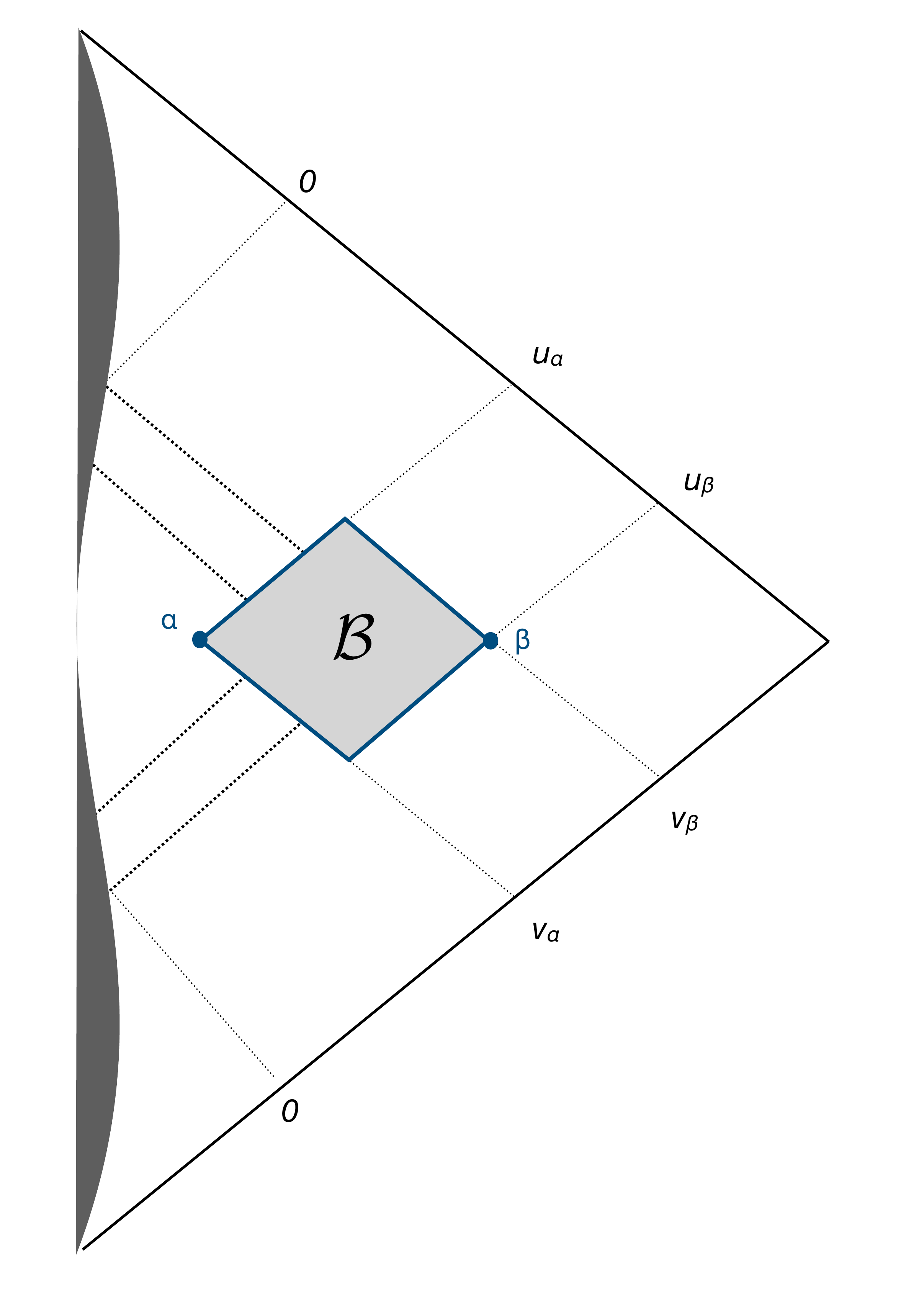}}
\caption{Conformal diagram of the black-to-white hole spacetime.}
\label{bouncestarfull3}
\end{center}
\end{figure}

With these definitions in place, we  now come to the key point of the construction. We excise from the spacetime the entire region surrounded by the blue line in Fig.~\ref{bouncestarB}.  We identify $\beta_L$ with $\beta_U$ and the $(t=0,r>r_\beta)$ surface in the lower asymptotic region with the  $(t=0,r>r_\beta)$ surface in the upper asymptotic region.  The gluing is  possible, since these are isometric surfaces with vanishing extrinsic curvature in the two isometric outer regions.   Call $\cal B$ the spacetime diamond defined by $\alpha$ and $\beta\equiv \beta_L=\beta_U$ and discard any previous information about the metric inside $\cal B$.   The resulting spacetime is the black-to-white hole spacetime we were looking for and it has the Penrose diagram depicted in Fig.~\ref{bouncestarfull3}. 

The geometry outside the $\cal B$ region depicted in Fig.~\ref{bouncestarfull3} is everywhere locally isomorphic to the geometry in the exterior of the blue lines depicted in Fig.~\ref{bouncestarB}, but the two are not globally isomorphic. The  interior region $S$ bounded by a timelike singularity discovered in the spacetime constructed in section~\ref{s:ext} is not present in the black-to-white hole spacetime. There is a unique asymptotic region in the exterior of both black and white hole.  As we shall see below, a non-singular metric can be assigned to the region $\cal B$. This will be done below, in Section \ref{effective}.

\section{Physical interpretation and large scale geometry}
\label{largescale}

Let us pause to discuss the physical interpretation and the logic of this construction and of the new parameters introduced. The advanced-time $v_\alpha$ is the time at which the horizon transition is triggered. The radial coordinate $r_{\alpha}$, which is uniquely specified by $v_\alpha$ and vice versa, is the maximum radius on the $t=constant$ surface in region $I$ containing the bounce point of the star for which the metric constructed in section~\ref{s:ext} is a good approximation of the spacetime of a black hole. The radial coordinate $r_{\sigma}$ is the maximum radius on the $v=v_\alpha$ surface for which the quantum physics of the horizons is non-negligible. The metric constructed in section~\ref{s:ext} is not a good approximation of the spacetime of a black hole in the future lightcone of $\sigma_L$, because it neglect the possibility of tunnelling. Since the black-to-white hole spacetime has a unique asymptotic region, the metric constructed in section~\ref{s:ext} must not be a good approximation of the spacetime of a real black hole also in the future of some surface reaching spacelike infinity in the lower region $L$. This is the $t=0$ surface identified by $\delta$ which intersect the outgoing component of the future lightcone of $\sigma_L$ in $\beta_L$. The radius $r_{\sigma}$ is completely specified once $r_{\alpha}$ and $r_{\beta}$ are given.

\begin{figure}
\begin{center}
\centerline{\includegraphics[width=5cm]{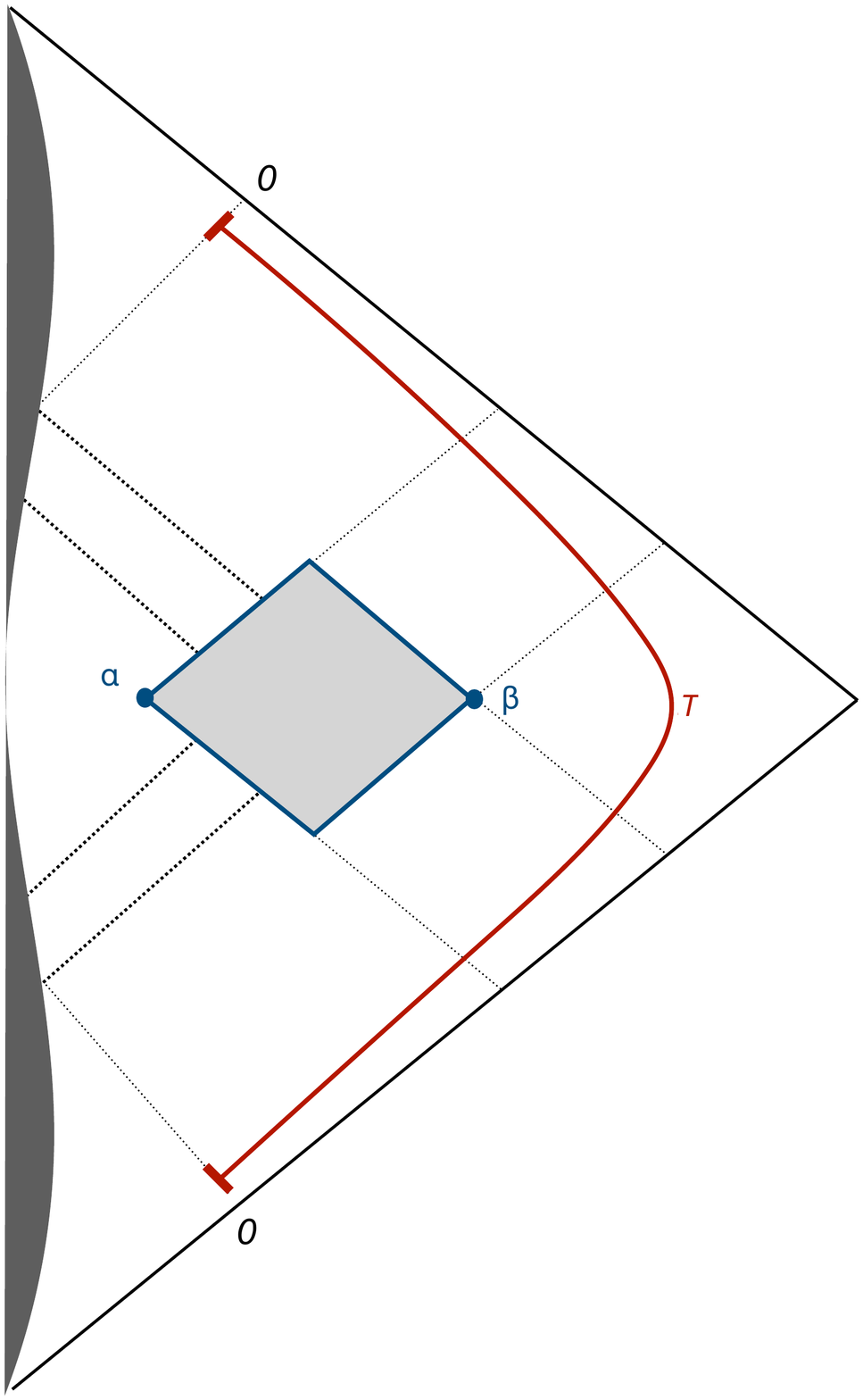}}
\caption{In red, the worldline of an observer moving at a constant distance $R\gg 2m$.}
\label{signals}
\end{center}
\end{figure}

Let's now consider the features of this geometry that can be measured at large radius.  At first sight,  since the geometry at a large distance from the hole is the Schwarzschild geometry, one might think that the only parameter measurable at large distance is the mass $m$, but this is wrong. 

Consider an observer that remains at distance $R\gg 2m$ from the hole. Consider their proper time $T$ between their $v=0$ advanced time and their $u=0$ retarded time (that is from the advanced time in which the star enters its horizon and the retarded time in which the star exits it). Their worldline is shown in red in Fig.~\ref{signals}. By symmetry, $T$ is twice the proper time along this worldline between the $v=0$ advanced time and the $t=0$ surface, namely the proper time of the worldline in red in Fig.~\ref{all}. This is approximately (minus) the $t$-coordinate $t_R$ of the observer at $v=0$, that is 
\begin{equation}
T/2\sim-t_R=r_*(R)-v=r_*(R).
\end{equation}
For $R\gg m\gg m_{Pl}$, recalling that we have fixed $r_*(r_\delta) = 0$, we have
\be
 r_*(R)\sim  R+2m\ln(R-2m)-2m \ln{\delta}. 
 \label{Rlogdelta}
\ee
Using this,
\begin{equation}
T\sim 2R+4m\ln(R-2m)-4m \ln{\delta}.
\label{Tdelta}
\end{equation}
The first two terms of this expression depends on $R$. Not so the last term  \be
{\cal T}\equiv - 4m \ln{\delta}. 
\label{tau}
\ee
 This is independent from the observer and \emph{is large and positive when $\delta$ is small}.  This means that $\delta$ can be measured by comparing the proper times of two distant observers.

Let's see this more explicitly, since it is a key point.  The first term in eq.~\eqref{Tdelta}, namely $2R$, is the travel-time of light from an observer at radius $R$ to the center and back, in flat spacetime.  The second (logarithmic) term is a relativistic correction to this travel time in the Schwarzschild geometry.  This can be seen by comparing $T$ with the corresponding proper time $T'$ of a second distant observer at a constant radius $R'$ satisfying $R\gg R' \gg 2m$.
The difference of these proper times is 
\begin{eqnarray}
T-T'&\sim& 2R-2R'+2m \ln(R-2m)- 2m\ln(R'-2m)
\nonumber \\
&\sim& R+2m\ln(R-2m),
\label{diffT}
\end{eqnarray}
which shows that the first two terms in eq.~\eqref{Tdelta} simply account for the back and forward travel-time of light and they are not related to the actual lifetime of the hole.

The quantity $\cal T$ is therefore a parameter that can be measured from a distance and characterises the intrinsic duration of the full process of formation of the black hole, tunnelling into a white hole and dissipation of the white hole. We can therefore properly \emph{call} the quantity $\cal T$ the duration of the bounce, or `bounce time'. We have thus found the geometrical interpretation of $\delta$ in terms of the total bounce time $\cal T$:
\be
\delta=e^{-\frac{\cal T}{4m}}.
\ee

Notice that $\delta$, unlike $r_\alpha$ and $r_\beta$ and in spite of being small, is a macroscopic parameter.   Namely it is a parameter of the global geometry that can be determined by measurements at large distance from the hole. The gluing of the upper and lower regions in Fig.~\ref{bouncestarB} introduces this global parameter, in the same manner in which gluing two portions of flat space can introduce the radius of a cylinder: a global parameter not determined by the local geometry.  The two other parameters $r_\alpha$ and $r_\beta$ determine only the location of the $\cal B$ region, without affecting the observations at large distance.  Large distance observations are therefore determined by two parameters only: the mass $m$ of the star and $\delta$, or the bounce time ${\cal T}=- 4m \ln{\delta}$.

\section{Global coordinates}
\label{globalcoordiates}

\begin{figure}
\begin{center}
\centerline{\includegraphics[width=5cm]{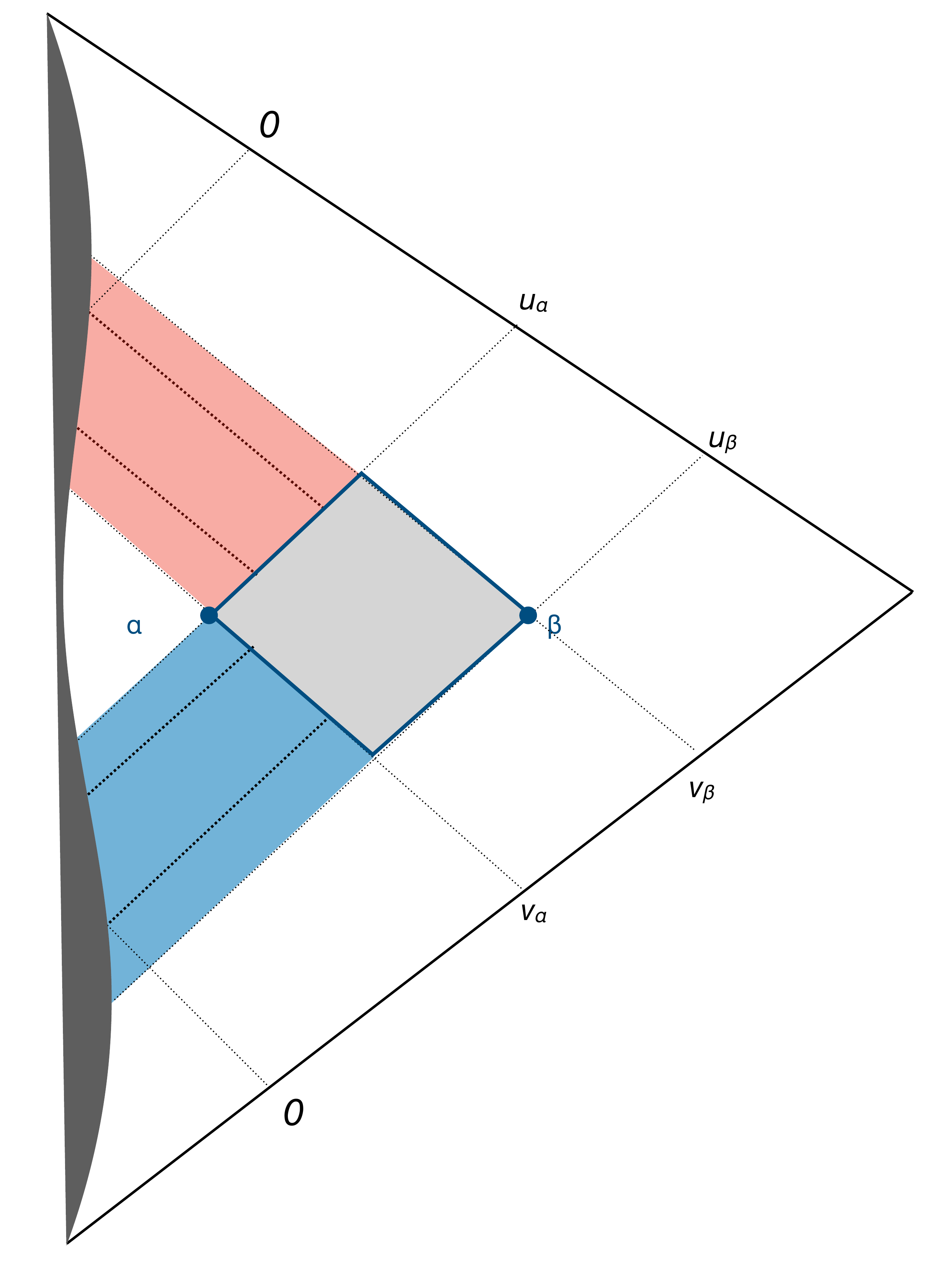}}
\caption{In red, the region defined by
$v\in[{v_\alpha},v_\beta]$ and $u\in[u_{star} (v),u_\alpha]$. In blue, the region defined by $u\in[{u_\alpha},u_\beta]$ and $v\in[v_{star} (u),v_\alpha]$.}
\label{regions22}
\end{center}
\end{figure}

 Using eqs.~(\ref{vu}-\ref{r*vu}) the $v$ coordinate can be defined everywhere except for the region specified by $v\in[{v_\alpha},v_\beta]$ and $u\in[u_{star} (v),u_\alpha]$, where $u_{star} (v)$ represent the wordline of the boundary of the star in $(u,v)$ coordinates. This region is depicted in red in Fig.~\ref{regions22}.   If we continue the $v$ coordinate into this red region, it diverges on the two horizons.    Similarly, the $u$ coordinate is well defined everywhere except for the region specified by $u\in[{u_\alpha},u_\beta]$ and $v\in[v_{star} (u),v_\alpha]$, represented in blue in Fig.~\ref{regions22}.

In this section we define well-behaved global coordinates outside region $\cal B$ (and outside the star). This will allow us to write a regular and singularity-free metric in region $\mathcal B$ in the next section.

\begin{figure}
\begin{center}
\centerline{
\includegraphics[width=6cm]{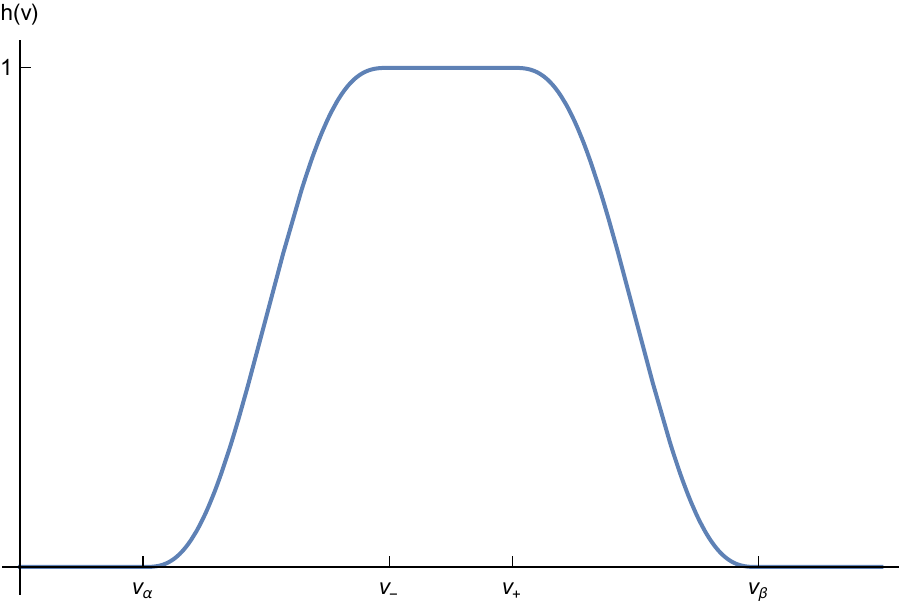}}
\caption{The interpolating function $h(v)$ defined in Footnote 1.}
\label{plot2}
\end{center}
\end{figure}

Starting from the coordinate $v$, introduce a smooth function $f(v)$ such that $f(v)=v$ for $v<{v_\alpha}$ and $v>v_\beta$, while for $v\in[{v_{\alpha}},v_\beta]$ the function $f(v)$ ranges in $[{v_\alpha},\infty]\cup[\infty,-\infty]\cup[-\infty,v_\beta]$, diverging logarithmically in two points, that we call $v_+$ and $v_-$.    Specifically, let
\be
f(v)=v+R(v)
\label{prima}
\ee
where $R(v)=0$ outside the interval $v\in [v_\alpha,v_\beta]$, and in this interval is defined as
\be
 R(v)=2h(v)\big({c_+}\log|v-v_+|+{c_-}\log|v-v_-|\big),
 \label{seconda}
\ee
with $v_\alpha<v_-<v_+<v_\beta$ and $c_\pm =1/F'(r_\pm)$. (The constants $c_\pm$ multiply the divergent logarithms in the expression of $r_*(r)$ in eq.~\eqref{r*exact}.) The function $h(v)$ can be chosen to be any function that interpolates smoothly between $h(v_\alpha)=h(v_\beta)=0$ and $h(v_-)=h(v_+)=1$, and has vanishing derivatives up to an arbitrary order $n$ in these four points\footnote{A simple example is $h(v)\!=\!0$ for $v\!<\!v_\alpha$ and  $v\!>\!v_\beta$,   $h(v)\!=\!1$ for $v\in[v_-,v_+]$,  $h(v)=S_n((v-v_\alpha)/(v_--v_\alpha))$ for   $h\in[v_\alpha,v_-]$, and  $h(v)=1-S_n((v-v_+)/(v_\beta-v_+))$ for $v\in[v_+,v_\beta]$, where $S_n(x)$ is the $n$-th order "smooth step" function that interpolates between $S_n(0)=0$ and $S_n(1)=1$, with vanishing derivatives up to order $n$ at $x=0$ and $x=1$ \cite{wiki}. For instance, $S_2(x)=6x^5-15x^4+10x^3$.}.  See Fig.~\ref{plot2}.

We then define a new $v$ coordinate in the red region by 
\be
f(v)=2r_*(r)-u\, ,
\label{newtransf}
\ee
instead than eq.~\eqref{vu}. The coordinate $v$ defined in this way covers the red region in its range $v\in[{v_\alpha},v_\beta]$ and  matches with the $v$ coordinate defined elsewhere.  Notice that $(2r_*(r)-u)$ diverges on the horizons,  but $v$, so defined, does not: on the horizons it takes the finite values $v_-$ and $v_+$.   Hence $u$ and (this newly defined) $v$ are finite continuous coordinates in the red region. For $v$ to be a good coordinate for the region, we also need to check that the metric is well-defined there. This can be done as follows. 

\begin{figure}[b]
\begin{center}
\centerline{
\includegraphics[width=7cm]{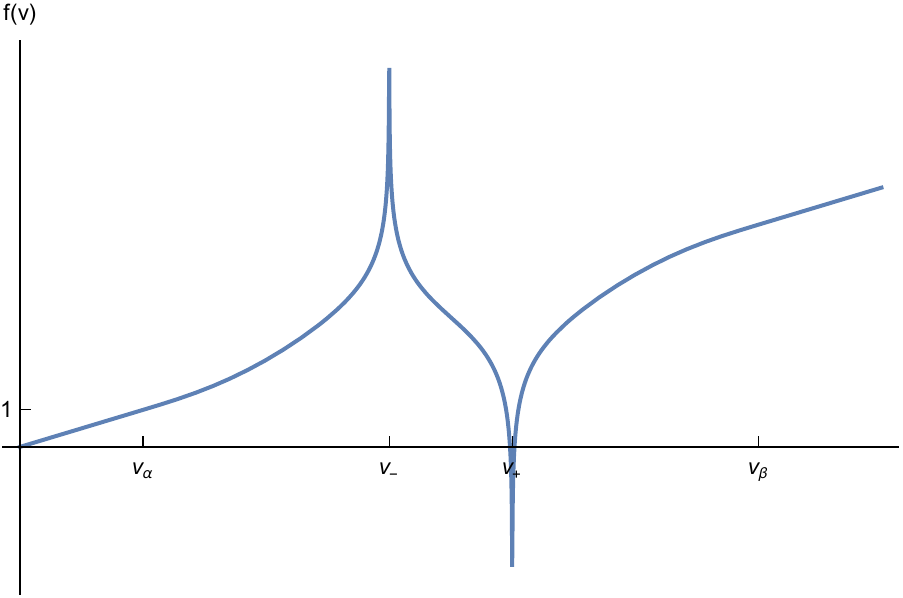}}
\caption{The function $f(v)$ defined in eqs.~(\ref{prima}-\ref{seconda}).}
\label{plot2b}
\end{center}
\end{figure}

The line element in the red region reads
\be
\diff s^2=F(r(u,v))f'(v)\,\diff u\,\diff v+r^2(u,v)\diff\Omega^2.
\label{V}
\ee
Near the horizon $r=r_\pm$ the function $F(r)$ has a zero of the form $r-r_\pm$ while $f'(v)$ diverges as the derivative of the logarithm, namely $1/(v-v_\pm)$. In particular, the $g_{uv}$ component of the metric behaves as
\be
g_{uv}=\frac{F(r)f'(v)}{2} \sim  \frac{r-r_\pm}{v-v_\pm}
\label{guv}
\ee
near the horizon $r=r_\pm$. Let us now study the transformation in eq.~\eqref{newtransf} around the horizons. For $r\sim r_\pm$,  eq.~\eqref{r*exact} gives
\be
r_*(r) \sim  c_\pm \log |r-r_\pm| + \mu_1\, ,
\ee
with 
\be
\begin{split}
\mu_1 = & r_\pm + c_\mp \log |r_\pm-r_\mp| + \frac{c_1}{2} \log(r^2_\pm + a r_\pm + b) \\
& + \frac{(2c_1 /c_2-a)}{\sqrt{b-a^2/4}} \tan^{-1} \left(\frac{r_\pm +a/2}{\sqrt{b-a^2/4}}\right)+ K\, .
\end{split}
\ee
If $v\sim v_\pm$, then eqs.~(\ref{prima}-\ref{seconda}) give
\be
f(v) \sim  2 c_\pm \log |v-v_\pm| + \mu_2\, ,
\ee
with 
\be
\mu_2 =   c_\mp \log |v_\pm-v_\mp|  .
\ee
This means that near the horizon $r=r_\pm$ eq.~\eqref{newtransf} reads
\be
2 c_\pm \log |v-v_\pm| + \mu_2 - 2c_\pm \log |r-r_\pm| - 2\mu_1 \sim u\, ,
\ee
namely
\be
\frac{r-r_\pm}{v-v_\pm} \sim e^{-\frac{2\mu_1 -\mu_2}{2c_\pm}} e^{-\frac{u}{2c_\pm}}\,.
\ee
The metric component $g_{uv}$, and so the complete metric, is thus well-behaved around the horizons.

The same construction can be performed  in the symmetric $blue$ region. Given the values $u_\pm\equiv v_\pm$, and remembering that $u_\alpha=v_\alpha$ and $u_\beta=v_\beta$ by construction, we define a new $u$ coordinate in the blue region by 
\be
f(u)=2r_*(r)-v\, ,
\label{newtransf2}
\ee
where the function $f$ is given in eqs.~(\ref{prima}-\ref{seconda}). The coordinate $u$ defined in this way covers the blue region in its range $u\in[{u_\alpha},u_\beta]$ and matches with the $u$ coordinate defined elsewhere. The line element in the blue region reads
\be
\diff s^2=F(r(u,v))f'(u)\,\diff u\,\diff v+r^2(u,v)\diff \Omega^2
\label{U}
\ee
and it is well-behaved everywhere. This completes the construction of a global coordinate chart for the black-to-white hole spacetime. 

Summarizing, the line element is 
\be
\diff s^2=g(u,v)\,\diff u\,\diff v+r^2(u,v)\diff \Omega^2.
\label{null2}
\ee
In the white regions of Fig.~\ref{regions22}, namely where 
\begin{eqnarray}
    u\in [u_\beta,+\infty),&& v\in [v_{star} (u),+\infty)\, ,\\
u\in [u_{star} (v),+\infty),&& v\in [v_\beta,+\infty)\, ,\\
u\in [u_{star} (v_\alpha ),u_\alpha ],&& v\in [v_{star} (u),v_\alpha]\,,
\end{eqnarray}
we have 
\be
g(u,v)=F(r(u,v))
\label{null222}
\ee
and the radius $r(u,v)$ is implicitly given by 
\be
2r_*(r)=v+u\,.
\label{r}
\ee
In the red region specified by
\begin{equation}
u\in[u_{star} (v),u_\alpha],\: v\in[{v_\alpha},v_\beta]\, ,
\end{equation}
we have 
\begin{equation}
g(u,v)=F(r(u,v))f'(v)    
\end{equation}
and the radius $r(u,v)$ is implicitly given by
\be
2r_*(r)=f(v)+u=v+u+R(v).
\label{43}
\ee
In the blue region specified by
\begin{equation}
u\in[{u_\alpha},u_\beta],\: v\in[v_{star} (u),v_\alpha]\, ,
\end{equation}
we have 
\begin{equation}
g(u,v)=F(r(u,v))f'(u)  
\end{equation}
and the radius $r(u,v)$ is implicitly  given by
\be
2r_*(r)=v+f(u)=v+u+R(u). 
\label{44}
\ee
This metric is well-behaved everywhere and, thanks to the interpolating function $h$ in eq.~\eqref{seconda}, it joins regularly (up to an arbitrary order $n$) at the boundaries of the red and blue regions.

\section{An effective metric in the \texorpdfstring{$\cal B$}{B} region}
\label{effective}

Can the $\cal B$ region be filled with an effective Lorentzian metric that joins regularly with the exterior metric at their boundary? To show that the answer is affirmative, let us now construct one such metric. 

\begin{figure}[b]
\begin{center}
\centerline{\includegraphics[width=7cm]{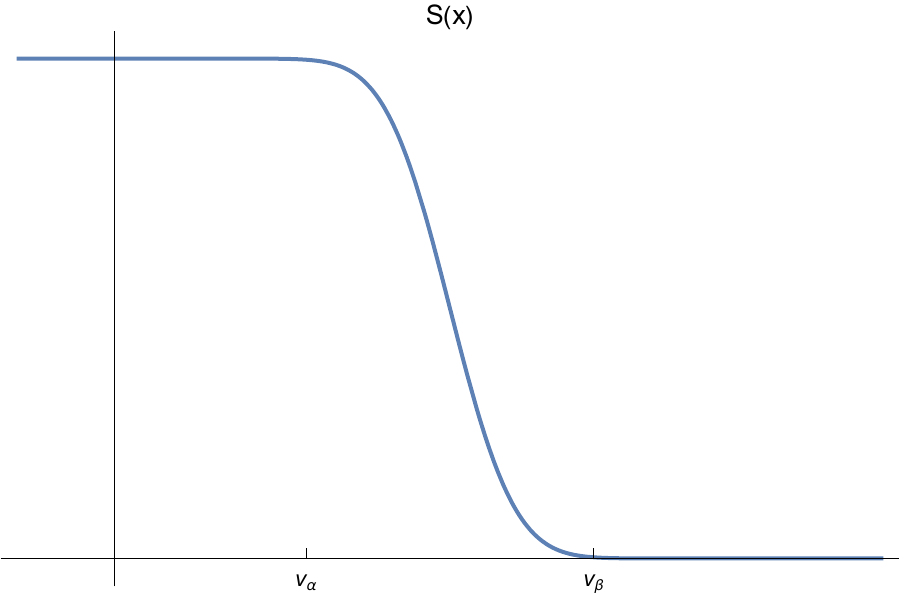}}
\caption{The interpolating function $S(x)$ defined in eq.~\eqref{S}}
\label{erf}
\end{center}
\end{figure}

We can write the metric constructed in the last section in a more compact form by choosing a regular-enough function $S(x)$ such that $S(x)=1$ for $x<{v_\alpha}$, $S(x)=0$ for $x>{v_\beta}$ and $S(x)$ interpolates between these two values in $x\in [v_\alpha,v_\beta]$. For instance,
\be
S(x)=1-S_n((x-v_\alpha)/(v_\beta-v_\alpha))
\label{S}
\ee
in $x\in [v_\alpha,v_\beta]$, where $S_n(x)$ is the $n$-th order smooth step function mentioned in Footnote 1. The function $S(x)$ is represented in Fig.~\ref{erf}.
This allows us to write compactly (see eq.~\eqref{null2})
\be
g(u,v)=F(r(u,v))\: f(u,v)\, ,
\label{metricin}
\ee
where
\be
f(u,v)=(1+S(u)R'(v))\;(1+S(v) R'(u))
\label{metricin2}
\ee
and $r(u,v)$ is implicitly defined by
\be
2r_*(r)=v+u+ S(u)R (v) + S(v) R(u)\,.
\label{rfinal}
\ee
The interpolating function $S(x)$, so far, serves only to simplify notation: it does not actually affect the metric, which for the moment does not regard the $\cal B$ region, defined by
\be
u\in[{u_\alpha},u_\beta],\: v\in[v_{\alpha},v_\beta]\, .
\ee

It is now easy to perform the standard conformal transformation $u=\tan U$, $v=\tan V$ to bring the coordinates in a finite and compact domain, but we do not do this explicitly. The coordinates $(U,V)$ are those in which all the Penrose diagrams of this article are drawn.

To extend the metric to the $\cal B$ region, the idea is to extend eqs.~\eqref{null2} and~(\ref{metricin}-\ref{rfinal}) to the $\cal B$ region.   The global coordinate system $(u,v)$ constructed in the last section extends naturally to this region, because the coordinate intervals are the same on the opposite sides of the diamond boundary of the $\cal B$ region.  Furthermore, thanks to the properties of the function $R$, it is easy to show that the functions $f(u,v)$ and $r(u,v)$ defined on the whole black-to-white hole spacetime (outside the star) joins regularly (up to an arbitrary order $n$) at the boundary of the region $\cal B$.

Eqs.~\eqref{null2} and~(\ref{metricin}-\ref{metricin2}) can then be used to extend the metric to the complete black-to-white hole spacetime outside the star, thus providing an (arbitrary) effective Lorentzian metric describing the interior of the region $\cal B$.

\section{Horizons}
\label{horizons}

\begin{figure}[tbp]
\begin{center}
\centerline{\includegraphics[width=5cm]{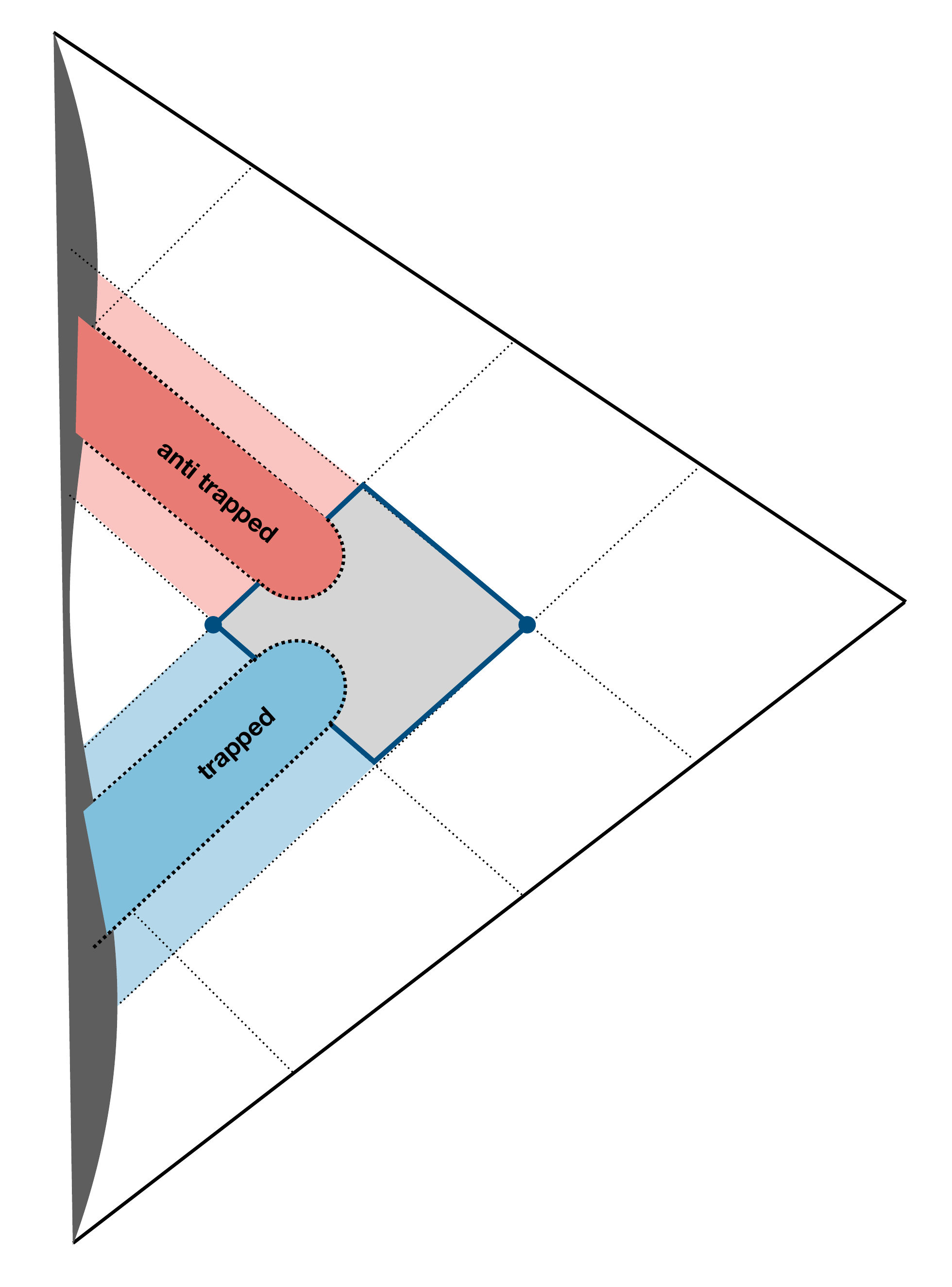}}
\caption{One of the possible qualitative behaviours of the apparent horizons.}
\label{apparentH}
\end{center}
\end{figure}

Finally, we study the structure of the horizons defined by the Lorentzian metric we have constructed in region $\cal B$. 

There are no event horizons: the past of future null infinity is the entire spacetime. 

There are no global killing horizons. This is due to the fact that the local Killing symmetry is broken in the $\cal B$ region (and in the star).   This can be shown as follows.  The norm $|\xi|$ of a Killing field $\xi$ is conserved along its own integral lines because the Lie derivative ${\cal L}_\xi |\xi|={\cal L}_\xi (g_{ab}\xi^a\xi^b)$ vanishes, as the Lie derivative of each factor does.   Take one of the killing horizons outside region $\cal B$, say $u=u_\pm$.  It is a null integral line of the Killing field.  If the Killing symmetry was respected in $\cal B$, its integral line would remain null. So, it would follow the null geodetic.   The null geodetic is $u=constant$, so the killing horizon would have to continue to the outer region through region $\cal B$. But it does not. Hence, the Killing symmetry is broken inside the $\cal B$ region and there is no global killing horizon\footnote{We thank Alejandro Perez for pointing this out.}.

This is comprehensible physically: what happens inside the $\cal B$ region is a quantum tunnelling, and a tunnelling breaks stationarity.  This, by the way, is why calculations that impose a global killing symmetry outside the star miss the possibility of the tunnelling. 

The horizons in the red and blue regions are however not only local killing horizons, but also apparent horizons.  That is, they separate trapped, non-trapped and anti-trapped regions. These regions can be characterized by the causal character of the $r=constant$ surfaces, which are timelike in the non-trapped regions and spacelike in the trapped and anti-trapped regions.  By continuity, the apparent horizons must continue inside the $\cal B$ region.  How?

\begin{figure}[t]
\begin{center}
\centerline{\includegraphics[width=5cm]{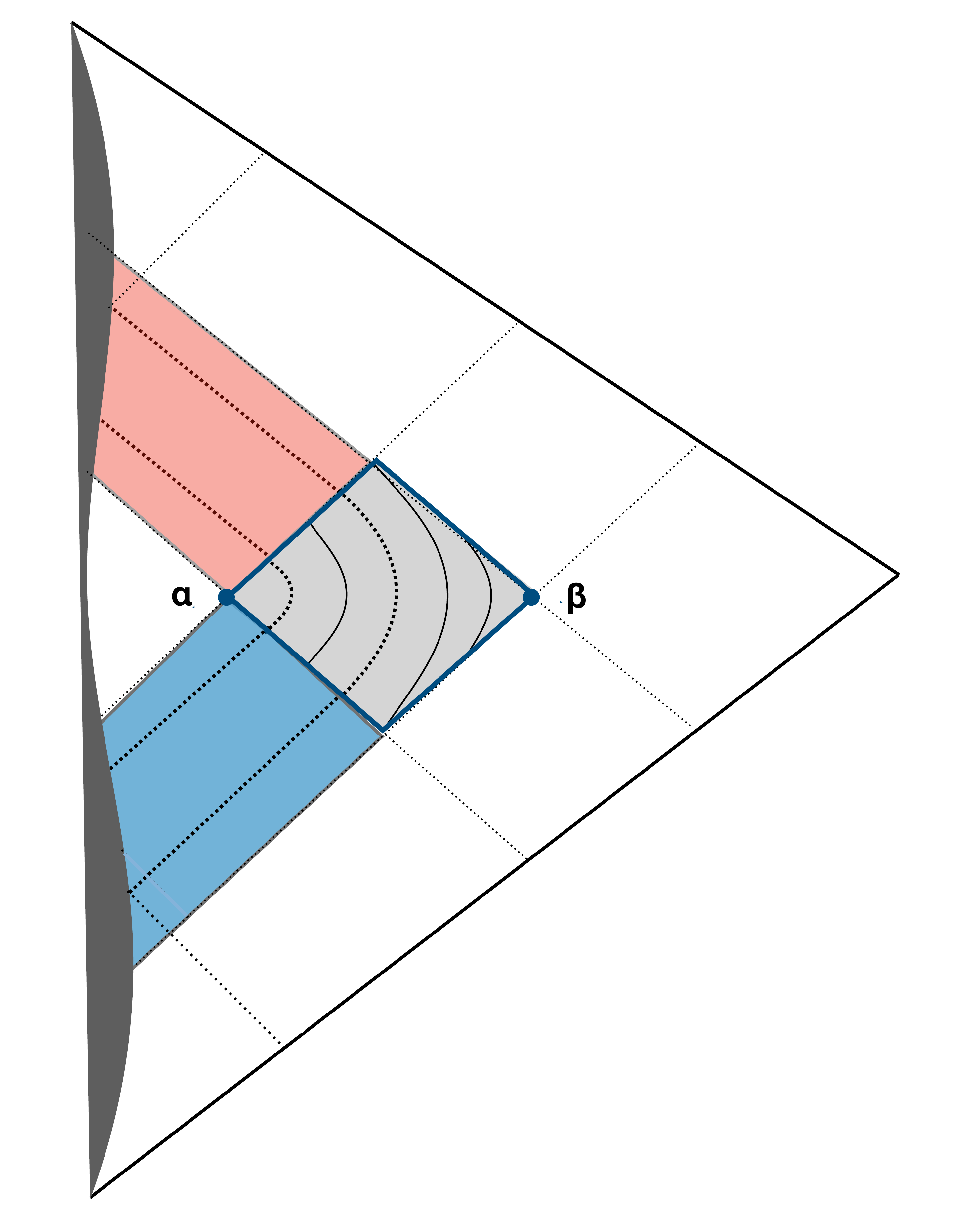}}
\caption{One of the possible qualitative behaviours of the surfaces of equal radius.}
\label{radius}
\end{center}
\end{figure}

The qualitative way they continue inside $\cal B$ follows from a topological consideration.   The overall spacetime is symmetric under a past$\leftrightarrow$future flip. Call $\Sigma_0$ the $u=v$ reflection surface.  By reflection symmetry, the $r=constant$ surfaces can only be either parallel or orthogonal to $\Sigma_0$.  Outside region $\cal B$ they are clearly orthogonal to $\Sigma_0$, both in the asymptotic exterior region and in the interior region where the star's bounce takes place.   By continuity, since the $r=constant$ surfaces cannot jump from orthogonal to parallel to $\Sigma_0$, they must be (almost) everywhere orthogonal to $\Sigma_0$, also inside region $\cal B$. Given that only timelike surfaces can be orthogonal to $\Sigma_0$, the internal non-trapped region is expected to be connected to the external one through the region $\cal B$. A possible way for this to happen is that the apparent horizons qualitatively behave as in Fig.~\ref{apparentH}, making sure that the trapped and anti-trapped regions are compact and do not share a finite boundary.  The surfaces of constant radius would then have the qualitative form represented in Fig.~\ref{radius}.

Other possible topological structures for the constant-radius surfaces and for the trapped and anti-trapped regions can result from different choices of the interpolating metric and in particular distinct relative values of the parameters $v_\alpha<v_-<v_+<v_\beta$. Given that the metric in $\mathcal{B}$ may be highly dynamical, there are possibly other compact trapped/anti-trapped regions created in $\mathcal{B}$ in addition to the ones shown in Fig.~\ref{apparentH}.

\section{Conclusions}

We have constructed a spacetime geometry that describes the collapse of a spherically symmetric pressure-less star, the subsequent formation of a black hole, the bounce of the star, the quantum transition of the black hole into a white hole and the final expansion of the star out of the white hole.   The entire geometry outside the star is given in a single global null coordinate patch.  The metric satisfies the Einstein field equations at a distance from the quantum transition region.  If the mass of the star is large compared to the Planck mass, this classical region includes a large portion of the interior of the black and the white holes. 

The geometry of the classical region is determined by two parameters: the mass $m$ of the star and the global duration $\cal T$ of the process, from the collapse of the star to its emersion from the white hole.  The duration $\cal T$ can be determined by measurements at large distance from the hole.  Since this duration is not determined by the initial conditions and the classical Einstein field equations, it must be determined (probabilistically) by the quantum theory as a function of $m$ and $\hbar$, like the lifetime in radioactive decay.   A quantum theory of gravity must provide the probability distribution of $\cal T$ as a function of $m$  \cite{Christodoulou2016, DAmbrosio2021, Soltani2021}.  In the classical limit, ${\cal T}\to\infty$ and black holes are eternal. 

The geometry of the full spacetime depends also on microscopic parameters relating to quantum gravity effects but not affecting the observation at large distance. Two of these parameters, $r_\alpha$ and $r_\beta$, determine the location and the size of the horizon tunneling region $\cal B$. These two parameters are not however the only microscopic parameters determining the geometry in $\cal B$. The latter depends also e.g. on the arbitrarily chosen interpolating function $S(v)$. Although this geometry depends on some choices, it is still remarkable that there exists a regular metric in $\cal B$, given that this region was a mystery in the earlier studies of the black-to-white hole transition.

The regular metric that interpolates the geometry within the horizon tunneling region that we have constructed is sensitive to short-distance quantum gravity effects. 
This is only a proof of existence: uniqueness is beyond the scope of this paper. It could be interesting to better understand the metric in this region in terms of effective equations that could fix the ambiguity.  Still, getting a sense of the size of this tunnelling region may be interesting. This can be done for instance by computing the length of the spacelike curve $u=v$ from $r_\alpha$ to $r_\beta$ and the proper time along the other formal diagonal of the diamond.  We leave this as an exercise to the reader.  On the other hand, the size and shape of the boundary of the region $\cal B$ are crucial for the quantum calculation of the transition amplitude \cite{Christodoulou2016, DAmbrosio2021, Soltani2021}. 

Although no dynamical equations are involved for constructing the geometry outside the star, the existence of the regular metric of the entire spacetime and, in particular, of region $\cal B$ suggests that certain effective dynamics of spherical symmetric quantum gravity should be able to derive the geometry from first principles (see e.g. \cite{Giesel2022,Han2022,Han:2022rsx,Ashtekar2018b,Ashtekar2018d,Husain2022,Olmedo2017,Husain2022,Gambini2020,Bodendorfer:2019cyv} for some recent progress on the effective dynamics of spherical symmetric black hole).

The metric we have constructed has much in common with the Reissner-Nordstr\"om and Kerr metrics, with the fundamental difference that it avoids all singularities of those geometries. See also \cite{Rignon-Bret2022}.  Importantly, it also avoids the Cauchy horizon instability of these metrics
\cite{simpson1973internal,dafermos2003stability,Carballo-Rubio2018a,Carballo-Rubio2021}: no observer crossing the inner horizon receives an infinitely blue-shifted energy from outside the hole.\footnote{We thank Cong Zhang for a discussion on this point. The issue is also discussed in \cite{Rignon-Bret2022}.}

We have neglected Hawking radiation under the assumption that its effects are negligible in a first approximation of the phenomenon. If we take the Hawking radiation into account the relevant mass for the phenomenon is not the initial mass $m_0$ of the star anymore, but the actual shrinking mass $m$ of the evaporating hole, determined by the horizon area, because the tunneling of the horizons is a phenomenon regarding the local geometry of the horizons.   In a realistic black hole, the accumulation of quantum effects trying to trigger the horizon transition and the Hawking evaporation happen at the same time. The geometry described here must then be corrected to account for the earlier evaporation phase and the fact that the size of the interior of the hole is determined by its age and not by the area of the horizon~\cite{Christodoulou2015,Christodoulou2016ab}. Therefore we expect only the tunnelling region of the geometry described here to be relevant for a realistic situation, not the long term evolution.  We nevertheless expect the two main large scale parameters to remain key observables at large distance in general. 

It is reasonable to expect that the closer is the shrinking mass $m$ to the Planckian value, the more probable is the horizon transition to be triggered. For a macroscopic black hole of initial mass $m_0$ it takes a time of the order $m_0^3$ for the mass $m$ to reach a Planckian value, and therefore for the probability of the transition to be of order unity. In this scenario the lifetime of the black hole would thus be long. Furthermore, the resulting white hole would be of Planckian size and it may not suffer the Eardley instability~\cite{Eardley:1974zz}, being stabilized by quantum gravity as discussed in~\cite{rovelli2018small}, opening an intriguing potential connection with dark matter.  This is possible because most of the energy of the black hole is emitted via the Hawking radiation before the horizon transition, while the information can remain trapped inside the hole and be emitted slowly during the long life of the white hole~\cite{Rovelli2017e,Bianchi2018e,Rovelli2019a,Kazemian2022}.

\begin{acknowledgments}
The authors thank Francesca Vidotto, Edward Wilson-Ewing, Viqar Husain, Hongguang Liu, Cong Zhang, Simone Speziale and Alejandro Perez for useful exchanges. A special thanks to the quantum gravity group at Western University, where this research was done. Western University is located in the traditional lands of Anishinaabek, Haudenosaunee, L\=unaap\`eewak, and Attawandaron peoples.
This research was made possible thanks to the project on the Quantum Information Structure of Spacetime (QISS) supported by the JFT grant 61466. 
M.H. receives support from the National Science Foundation through grants PHY-1912278 and PHY-2207763, and the sponsorship provided by the Alexander von Humboldt Foundation. In addition, M.H. acknowledges IQG at FAU Erlangen-N\"urnberg, IGC at Penn State University, Perimeter Institute for Theoretical Institute, and University of Western Ontario for the hospitality during his visits. 
C.R. acknowledges support by the Perimeter Institute for Theoretical Physics  through its distinguished research chair program. Research at Perimeter Institute is supported by the Government of Canada through
Industry Canada and by the Province of Ontario through the Ministry of Economic Development and Innovation.
F.S.'s work at Western University is supported by the Natural Science and Engineering Council of Canada (NSERC) through the Discovery Grant "Loop Quantum Gravity: from Computation to Phenomenology".
\end{acknowledgments}

\section*{Appendix A: Zeros of \texorpdfstring{$F(r)$}{F(r)}}
We want to find the zeros of the function
\be
F(r)=1-\frac{2m}{r}+\frac{A m^2 }{r^4}
\ee
with $A$ being a constant with dimensions of a squared mass and satisfying $A \ll m^2$. Finding the zeros of $F(r)$ is equivalent to finding the roots of the fourth-degree equation
\be
r^4 -2mr^3+ A m^2 =0\,.
\label{eqq}
\ee
Although the exact solutions to this problem are known, their expression is too complicated to be of any help in our analysis. Instead, we want to study these solutions perturbatively in the small parameter $A$.

To rigorously treat eq.~\eqref{eqq} as a perturbation problem in a small dimensionless parameter, let $x=r/m$, such that the equation to solve becomes
\be
x^4 -2x^3+ \varepsilon=0\,,
\label{problem}
\ee
where $\varepsilon := A/m^2 \ll 1$.
The unperturbed equation
\be
x^4 -2x^3=0\,
\ee
has the four solutions
\be
x_{1,2,3}=0 \quad\quad x_4=2\,.
\ee
We want to perturbatively search for solutions of eq.~\eqref{problem} of the form
\be
x_i=\sum_{n=0}^{\infty} a_{i,n} \varepsilon^n\, ,
\label{ansatz}
\ee
where $i=1,2,3,4$ and $a_{4,0}=2$, $a_{j,0}=0$ for $j=1,2,3$. The coefficients $a_{i,n}$ can be determined by solving eq.~\eqref{problem} order by order.

Let's start with the $\varepsilon$ order for $x_4$. Inserting 
\be
x_4 =2+a_{4,1}\varepsilon+ O(\varepsilon^2)
\ee
in eq.~\eqref{problem} we find
\be
\big(2+a_{4,1}\varepsilon+ O(\varepsilon^2)\big)^4 -2\big(2+a_{4,1}\varepsilon+ O(\varepsilon^2)\big)^3+ \varepsilon=0\,.
\ee
Solving to order $\varepsilon$ we obtain $a_{4,1}=-1/8$. This means that
\be
x_4= 2 -\frac{\varepsilon}{8}+O(\varepsilon^2)\,.
\ee
If we try to do the same for
\be
x_j =a_{j,1}+ O(\varepsilon^2),
\ee
where $j=1,2,3$, we get
\be
\big(a_{j,1}\varepsilon+ O(\varepsilon^2)\big)^4 -2 \big(a_{j,1}\varepsilon+ O(\varepsilon^2)\big)^3  + \varepsilon=0\,.
\ee
This equation is clearly not consistent, which means that the ansatz in eq.~\eqref{ansatz} is not consistent. It simply means that $x_j \sim \varepsilon$ ($j=1,2,3$) for $\varepsilon \ll 1$ is not true. In order to find the right scaling we can study the dominate balance of eq.~\eqref{problem} when $\varepsilon \ll 1$ (see \cite{bender}):
\begin{itemize}
\item[•] If $x^4 \sim x^3$, and thus 
	\be
	\varepsilon \ll x^4,x^3\,,
	\label{conscheck1}
	\ee
	we find one solution s.t. $x\sim 1$. Eq.~\eqref{conscheck1} gives $\varepsilon \ll 1$, which is consistent. This solution is the solution 	$x_4$ we already found.

\item[•] If $x^4 \sim \varepsilon $, and thus 
	\be
	x^3 \ll x^4,\varepsilon\,,
	\label{conscheck2}
	\ee
	we find three solutions s.t. $x\sim \varepsilon^{1/4}$. Eq.~\eqref{conscheck2} gives $\varepsilon^{3/4} \ll \varepsilon$, which is not consistent.
	
	\item[•] If $x^3 \sim \varepsilon $, and thus 
	\be
	x^4 \ll x^3,\varepsilon\,,
	\label{conscheck3}
	\ee
	we find three solutions s.t. $x\sim \varepsilon^{1/3}$. Eq.~\eqref{conscheck3} gives $\varepsilon^{4/3} \ll \varepsilon$, which is  consistent. Hence, the remaining solutions $x_j$ ($j=1,2,3$) behave as $x_j \sim \varepsilon^{1/3}$ for $\varepsilon \rightarrow 0$.

\end{itemize}

The new ansatz for the solutions $x_j$ ($j=1,2,3$) is then
\be
x_j=\sum_{n=1}^{\infty} b_{j,n} (\varepsilon^{1/3})^n\, .
\label{ansatz2}
\ee
Inserting 
\be
x_j=b_{j,1} \varepsilon^{1/3} + O(\varepsilon^{2/3})
\ee
in eq.~\eqref{problem} we find
\be
\big(b_{j,1}\varepsilon^{1/3} + O(\varepsilon^{2/3})\big)^4 -2 \big(b_{j,1}\varepsilon^{1/3}+ O(\varepsilon^{2/3})\big)^3  + \varepsilon=0.
\ee
Keeping only the order $\varepsilon$ we get $b_{j,1}^3=1/2$. The three solutions are thus
\be
b_{3,1}=\frac{1}{2^{1/3}} \quad\mathrm{and}\quad b_{(1,2),1}=\frac{1}{2^{1/3}} e^{\pm  2\pi i/3}\,.
\ee
All the subsequent orders of the solutions can be found in this way. 

The roots of eq.~\eqref{problem} to their second non-vanishing order in $\varepsilon$ are 
\be
\begin{split}
 x_{1,2}=& \left(\frac{\varepsilon}{2}\right)^{1/3} e^{\pm  2\pi i/3} + \frac{1}{6} \left(\frac{\varepsilon}{2}\right)^{2/3} e^{\pm  4\pi i/3} + O(\varepsilon)\, , \\
  x_3 =& \left(\frac{\varepsilon}{2}\right)^{1/3} + \frac{1}{6} \left(\frac{\varepsilon}{2}\right)^{2/3}+ O(\varepsilon)\, , \\
  x_4 =& \,2- \frac{\varepsilon}{8} + O(\varepsilon^2)\, .
\end{split}
\ee
Going back to the original variable $r$, the solutions to eq.~\eqref{eqq} to their second non-vanishing order in $A$ are
\be
\begin{split}
 r_{1,2}=& \left(\frac{A m}{2}\right)^{1/3} e^{\pm  2\pi i/3} + \frac{1}{6} \left(\frac{A}{2 \sqrt{m}}\right)^{2/3} e^{\pm  4\pi i/3}  \\
     & + O(A/m)\, , \medskip \\
  r_- = & \: r_3 = \left(\frac{A m}{2}\right)^{1/3} + \frac{1}{6} \left(\frac{A}{2 \sqrt{m}}\right)^{2/3}\\
    & + O(A/m) \, ,\medskip \\
  r_+ = & \: r_4 =  \,2m - \frac{A}{8 m} + O(A^2/m^3)\, .
\end{split}
\label{solperturb}
\ee

\begin{widetext}
\section*{Appendix B: The generalized tortoise coordinate \texorpdfstring{$r_*$}{r*}}

The generalized tortoise coordinate $r_*$ was defined in eq.~\eqref{r*r} as the coordinate satisfying
\be
\diff r_*=\frac{\diff r}{F(r)}.
\label{tortoise}
\ee
Let us integrate this differential equation. First of all, consider the fourth-degree equation
\be
r^4 -2mr^3+ A m^2 =0\,.
\label{eqq2}
\ee
The analysis in Appendix A tells us that this equation has two real solutions $r_\pm$ and two complex conjugate solutions $r_{1,2}$. This means that the polynomial $r^4 -2mr^3+ A m^2 $ can be rewritten as
\be
r^4 -2mr^3+A m^2 = (r-r_+)(r-r_-)(r^2+ a r + b)\, ,
\label{poly}
\ee
where $r^2+ a r + b=(r-r_1)(r-r_2)$ is a positive-definite second-degree polynomial. The values of $a$ and $b$ can be easily computed by expanding the polynomial in the right-hand side of eq.~\eqref{poly} and then equating it order-by-order to the left-hand side. This gives 
\be
a= (r_+ + r_-) -2m
\label{a}
\ee
and
\be
b= \frac{A m^2}{r_+ r_-}\,.
\label{b}
\ee
Eq.~\eqref{tortoise} can then be integrated as
\be
\begin{split}
r_* (r) = & \int \frac{\diff r}{1-2m/r+Am^2/r^4} = 
\int \frac{r^4 \,\diff r}{r^4-2m r^3+Am^2}
=  \int \diff r + \int \frac{2m r^3 - A m^2}{r^4-2m r^3+Am^2} \diff r  \\
= & \: r + \int \frac{2m r^3 - A m^2}{(r-r_+)(r-r_-)(r^2+ a r + b)} \diff r \, .\\
\end{split}
\ee
Using partial fraction decomposition we look for an expansion of the form
\be
\frac{2m r^3 - A m^2}{(r-r_+)(r-r_-)(r^2+ a r + b)} = \frac{c_+}{r-r_+} + \frac{c_-}{r-r_-} + \frac{c_1 r +c_2}{r^2+ a r + b}\, ,
\ee
where $c_+$, $c_-$ and $c_{1,2}$ are constants whose value need to be determined. By rewriting the right-hand side of this expression using a common denominator and then equating order-by-order the polynomials in the numerator of respectively left and right hand side we find
\be
c_+ = \frac{2 m r_+^3-A m^2}{(r_+ -r_-) \left(r_+^2+ a r_++b\right)}
= \frac{r_+^4}{(r_+ -r_-) \left(r_+^2+ a r_++b\right)} = \frac{1}{F'(r_+)}\,,
\label{cp}
\ee
\be
c_-= \frac{2 m r_-^3-A m^2}{(r_--r_+) \left(r_-^2 + a r_-+b\right)}= \frac{r_-^4}{(r_- -r_+) \left(r_-^2+ a r_-+b\right)} = \frac{1}{F'(r_-)}\,,
\label{cm}
\ee
\be
c_1 = -\frac{-2 a^2 m r_- r_++a A m^2-2 a b m r_--2 a b m r_++A m^2 r_-+A m^2 r_+-2 b^2 m+2 b m r_- r_+}{\left(r_+^2+ a r_++b\right) \left(r_-^2+ a r_-+b\right)}\,,
\label{c1}
\ee
\be
c_2 = -\frac{a^2 A m^2+a A m^2 r_-+a A m^2 r_+-2 a b m r_- r_+-A b m^2+A m^2 r_- r_+-2 b^2 m r_--2 b^2 m r_+}{\left(r_+^2+ a r_++b\right) \left(r_-^2+ a r_-+b\right)}\,.
\label{c2}
\ee
This leads to
\be
\begin{split}
r_* (r) = & \: r + c_+ \int \frac{\diff r}{r-r_+} + c_- \int \frac{\diff r}{r-r_-} + c_1 \int \frac{r + c_1 /c_2}{r^2+ a r + b} \diff r\\
= & \: r + c_+ \log |r-r_+| + c_- \log |r-r_-| + \frac{c_1}{2} \int \frac{(2r +a) + (2c_1 /c_2-a)}{r^2+ a r + b} \diff r\\
= & \: r + c_+ \log |r-r_+| + c_- \log |r-r_-| + \frac{c_1}{2} \log(r^2+ a r + b) + (2c_1 /c_2-a) \int \frac{\diff r}{(r+a/2)^2+ (b-a^2/4)}\\
= & \: r + c_+ \log |r-r_+| + c_- \log |r-r_-| + \frac{c_1}{2} \log(r^2+ a r + b) + \frac{(2c_1 /c_2-a)}{\sqrt{b-a^2/4}} \tan^{-1} \left(\frac{r+a/2}{\sqrt{b-a^2/4}}\right)+ K\,.
\end{split}
\label{r*exact}
\ee
The integration constant $K$ plays a key role: the fact that it can be independently fixed in two different regions that end up glued together is the technical reason for the appearance of the global geometrical parameter $\mathcal T$ (see eq.~\eqref{tau}) measuring the overall duration of the process described.

More precisely: we have picked an integration constant by posing $r_*(r_\delta)=0$. The constant $\delta$, which determines $\mathcal T$, is determined by the choice of the reflection surface, namely the gluing of the Lower and Upper regions. Formally, the choice of the reflection surface is equivalent to choosing the overlap between the lower $(v,r)$ coordinates and the upper $(u,r)$ coordinates. This is given by identifying $t=0$, namely, (from  $v=t+r_*(r)$ and  $u=-t+r_*(r)$) having $v=-u+2r_*(r)$.  Hence it is $r_*(r)$ that  determines which surface in $L$ we glue with which surface in $U$.   If we look only at the metric at large radius (for all times), we do not understand where the parameter $\mathcal T$ comes from.   It comes from the gluing, and the gluing is formally determined by the choice of the constant in $r_*(r)$.

\end{widetext}

\bibliographystyle{utcaps.bst}
\bibliography{ref.bib}

\end{document}